\documentclass[12pt]{article}

\usepackage{amssymb}
\usepackage{amsmath}
\usepackage{amscd}
\usepackage{latexsym}
\usepackage{graphicx}

\usepackage{enumitem}

\usepackage{cite}
\numberwithin{equation}{section}

\newcommand{\be}{\begin{equation}}
\newcommand{\ee}{\end{equation}}
\newcommand{\Dlt}{\Delta}
\newcommand{\dlt}{\delta}
\newcommand{\prt}{\partial}
\newcommand{\br}{{\bf r}}

\newcommand{\ba}{{\bf a}}

\newcommand{\bt}{\beta}

\newcommand{\ep}{\varepsilon}
\newcommand{\al}{\alpha}
\newcommand{\ra}{\rightarrow}
\newcommand{\sgm}{\sigma}

\newcommand{\gm}{\gamma}

\newcommand{\lbd}{\lambda}

\newcommand{\rgl}{\rangle}
\newcommand{\lgl}{\langle}

\newcommand{\cD}{{\cal D}}

\begin{document}

\begin{center}

{\Large{\bf Selected Topics of Social Physics: \\
Equilibrium Systems} \\ [5mm]

Vyacheslav I. Yukalov}  \\ [3mm]

{\it
$^1$Bogolubov Laboratory of Theoretical Physics, \\
Joint Institute for Nuclear Research, Dubna 141980, Russia \\ [2mm]

$^2$Instituto de Fisica de S\~ao Carlos, Universidade de S\~ao Paulo, \\
CP 369, S\~ao Carlos 13560-970, S\~ao Paulo, Brazil} \\ [3mm]

{\bf E-mail}: {\it yukalov@theor.jinr.ru}

\end{center}

\vskip 1cm

\begin{abstract}

The present review is based on the lectures that the author had been giving 
during several years at the Swiss Federal Institute of Technology in Z\"{u}rich 
(ETH Z\"urich). Being bounded by lecture frames, the selection of the material,
by necessity, is limited and is motivated by the author's research interests. 
The paper gives an introduction to the physics of social systems, providing 
the main definitions and notions used in the modeling of these systems. The 
behavior of social systems is illustrated by several simple typical models. 
The present part considers equilibrium systems. Nonequilibrium systems will be 
presented in the second part of the lectures. The style of the paper combines the 
features of a tutorial and a survey, which, from one side, makes it easy to read 
for nonspecialists aiming at grasping the basics of social physics, and from the 
other side, describes several rather recent original models containing new ideas 
that could be of interest to experienced researchers in the field.         

\end{abstract}

\vskip 3mm
 
{\bf Keywords}: social systems; typical agents; information and statistics; 
social phase transitions; yes-no model; regulation cost; fluctuating groups; 
self-organized disorder; population coexistence; country disintegration; 
formation of collective decisions
 
\newpage

{\large{\bf Contents}}

\vskip 5mm
{\bf 1. Introduction}

\vskip 3mm
{\bf 2. Principle of Minimal Information}

\vskip 2mm
   2.1. Information Entropy

   2.2. Information Functional

   2.3. Representative Ensembles

   2.4. Arrow of Time

\vskip 3mm
{\bf 3. Equilibrium Social Systems}

\vskip 2mm
   3.1. Free Energy

   3.2. Society Stability

   3.3. Practical Approaches

   3.4. Society Transitions

   3.5. Yes-No Model

   3.6. Enforced Ordering

   3.7. Command Economy

   3.8. Regulation Cost

   3.9. Fluctuating Groups

   3.10. Self-Organized Disorder

   3.11. Coexistence of Populations

   3.12. Forced Coexistence

\vskip 3mm
{\bf 4. Collective Decisions in a Society}

\vskip 2mm
   4.1. General Overview

   4.2. Utility Function

   4.3. Expected Utility

   4.4. Time Preference

   4.5. Stochastic Utility

   4.6. Affective Decisions

   4.7. Wisdom of Crowds

   4.8. Herding Effect

\vskip 3mm
{\bf 5. Conclusion}

\newpage

\section{Introduction}

Nowadays physics methods and models are widely used in social science for 
characterizing the behavior of different social systems. It is useful to stress 
that physics provides methods and models for describing social systems, but it 
does not pretend to replace social science. 

Under a social system one implies a bound collective of a number of interacting 
agents. These can be various human societies, starting with families, professional 
groups, members of organizations, financial markets, country populations etc. Or 
these can be animal societies forming biological systems, e.g. wolf packs, fish 
shoals, bird flocks, horse herds, swarms of bees, ant colonies, and like that. 
Thus the number of agents in a society is more than one and usually it is much 
larger than one ($N\gg 1$).

The collective of agents in a society is bound in the sense of having some 
features uniting the society members. For example, the members of a society can 
participate in joint activity, or can share similar goals or beliefs. The common 
features uniting the members make the collective stable or metastable, that is, 
bound for sufficiently long time that is much larger than the interaction time 
between the members. The agent interactions can be physical, economic, financial, 
and so on. Societies can be formed by force, as in the army and in prison, or 
can be self-organized.

Social systems pertain to the class of complex systems. A system is complex 
if some of its properties are qualitatively different from the agglomerated 
properties of its parts. Complex systems can be structured, consisting of parts 
that are complex systems themselves. For instance, the human world is composed 
of countries that, clearly, also are complex systems. In that sense, social 
systems usually are hierarchical, containing several levels of complex systems. 
For example, the human world is made up of countries that are composed of social 
groups that form organizations that include individuals whose bodies are made 
of biological cells and whose brains making decisions are extremely complex 
systems.

To represent a complex system, the society needs to have the following properties. 
The number of agents, composing the society, is to be large, $N\gg 1$, although 
this is not sufficient. If the members of the society are independent, so that the 
overall system can be characterized by a set of independent agents, this does not 
make such a system complex. In a complex society, its features are not just a sum 
of features of separate members. Thus bees can fly, but a bee colony is not just 
a swarm of flying bees. A group of people walking separately from each other is 
not a social system. A society becomes complex due to its agents interactions and 
mutual relations, which cause the nonadditivity of the society features. Strictly 
speaking, there is no a generally accepted mathematical measure of social system
complexity \cite{Waldrop_41,Bar_1997,Mitchell_42,Thurner_155,Ladyman_43}. Complexity 
is rather a qualitative notion assuming that: A society is complex if it is stable, 
consists of many interacting agents, so that the typical society features are not 
just an arithmetic average of features of separate individuals.

For example, an ant colony is a complex system, since the behavior of separate 
ants is regulated by their interactions and distribution of jobs: thus ant queens 
lay eggs, worker ants form an ant-hill, feed the queen, winged males mate with 
queens and die. Another example of a complex system is a bee colony, where a 
queen produces eggs, workers clean out the cells, remove debris, feed the brood, 
care for the queen, build combs, guard the entrance, ventilate the hive, forge 
for nectar, pollen, propolis, and water, and drones fertilize the queen and die 
upon mating. 

There are three interconnected problems in the description of complex social 
systems: how to model a society, how to investigate the model, and how it would 
be possible to regulate the society behavior. A model is to be simple, and at 
the same time realistic. Too simple models may not describe reality, but too 
complicated models can distort reality because of accumulation of errors in an 
excessively complicated description. Sometimes it happens that ``less is more" 
\cite{Browning_1}. So, it is desirable that models be not trivial, although 
not overcomplicated. 

In this review, first, the basic ideas are described allowing one to understand 
the general principles of constructing the society models, so that, from one 
side, the model would not be overcomplicated and, from another side, could catch 
the characteristic features of the considered society. Second, the main methods 
of investigating the system behavior are studied. Third, conditions are discussed 
making it possible to find the desired properties of the society and the related 
optimal parameters of models. 

The material of the article is based on the lectures that the author had been 
giving for several years at the Swiss Federal Institute of Technology in Z\"{u}rich
(ETH Z\"{u}rich). Overall, the content constitutes an one-year course consisting 
of two parts, one devoted to equilibrium (or quasi-equilibrium) systems, and the 
other, to nonequilibrium systems. Following this natural separation, the presentation
of the material is also split into the corresponding parts. The present part deals
with equilibrium social systems. The next part will consider nonequilibrium social
systems \cite{Yukalov_2023}.  
     
Being bounded by the lecture frames, the content of the review, by necessity, is
limited. The choice of the presented material is based on the researh interests
of the author.  

The layout of the present part is as follows. In Sec. 2, the principle of minimal 
information is discussed playing the pivotal role in establishing probability 
distributions for equilibrium and quasi-equilibrium systems. Section 3 studies some
simple typical models of equilibrium social systems. Section 4 gives the basics of
collective decision making in a society. Finally, Sec. 5 concludes.

\section{Principle of Minimal Information}

Practically all models, intending to describe the behavior of equilibrium or 
quasi-equilibrium complex systems, start with some extremization principles. 
These principles are widely spread in science as well as in life. Just as a 
joke, we may say that all life follows the extremization principle that can 
be formulated as ``minimum of labor and maximum of pleasure".    
 
To model social systems in complicated situations, when not all information is 
available, it is customary to resort to probabilistic description, defining the 
probability distribution from the principle of minimal information. This principle 
helps to develop an optimal description of a social system having limited amount 
of information on its properties. The information is always limited, since there 
are many agents and, in addition, their actions often are not absolutely rational. 
Experimental studies of brain activity clearly demonstrate that only a finite 
amount of information can be successfully processed by alive beings 
\cite{Loewenstein_2}. The rationality of social individuals is always bounded 
\cite{Simon_3}. Also, there exists random influence of environment. Moreover, 
not all information is even necessary for correct description, but excessive 
information can lead to the accumulation of errors and incorrect conclusions.

\subsection{Information Entropy}

Let us consider a social system of $N$ agents enumerated by the index 
$j=1,2,\ldots,N$. Generally, the number of agents can depend on time. Each $j$-th 
agent is marked by a set of characteristics
\be
\label{2.1}
 \sgm_j = \{ \sgm_{j\al} :~ \al = 1,2,\ldots, M_j \} \;  .
\ee  
The collection of the characteristics of all agents makes up the 
{\it society configuration set}
\be
\label{2.2}
X = \{ \sgm_j :~ j = 1,2,\ldots, N \} \;    .
\ee
The probability distribution $\rho(\sgm)$ over the variables $\sigma \in X$ is 
to be normalized,
\be
\label{2.3}
\sum_\sgm \rho(\sgm) = 1 \; , \qquad 0 \leq \rho(\sgm) \leq 1 \; ,
\ee
where the sum over $\sigma$ implies
$$
\sum_\sgm = \sum_{j=1}^N \; \sum_{\al=1}^{M_j} \;   .
$$

If all variables are equiprobable, then the distribution is uniform,
$$
\rho(\sgm) = \frac{1}{N_{tot}}  \qquad 
\left( N_{tot} = \sum_{j=1}^N M_j \right) \;   .
$$

A {\it statistical social system} is the triple
\be
\label{2.4}
 \{ N , ~ X, ~ \rho(\sgm) \} \;  .
\ee
  
However, the probability distribution needs to be defined. For this purpose, 
one introduces the information entropy $S$ that is a measure of the system 
uncertainty. It is a measure of unaccessible information, being a functional 
of the distribution over the society characteristics,
\be
\label{2.5}
S = S[\; \rho(\sgm)\; ] 
\ee
enjoying the following properties.

\vskip 2mm

(i) {\it Continuity}. It is a continuous functional of the distribution, such 
that for a small variation of the latter, the entropy variation is small:
\be
\label{2.6}
S[\; \rho(\sgm) + \dlt \rho(\sgm) \; ] - S[\; \rho(\sgm)\; ] \simeq
\frac{\dlt S[\; \rho(\sgm)\; ]}{\dlt\rho(\sgm)} \;
\dlt\rho(\sgm) \; .
\ee

\vskip 2mm
(ii) {\it Monotonicity}. For a uniform distribution,
\be
\label{2.7}
 S \left[\; \frac{1}{N_{tot}} \right]\;  <  \;
S \left[\; \frac{1}{N_{tot}+1} \right]\;  .
\ee

\vskip 2mm
(iii) {\it Additivity}. If a statistical system with the set of variables $\sgm$
can be separated into two mutually independent subsystems with the variable sets
$\sgm_1$ and $\sgm_2$, such that
$$
\rho(\sgm) = \rho(\sgm_1) \; \rho(\sgm_2) 
$$
and
$$
\sum_{\sgm_1} \rho(\sgm_1) = 1 \; , \qquad
\sum_{\sgm_2} \rho(\sgm_2) = 1 \;   ,
$$
then the entropy of the total system is a sum of the subsystems entropies,
\be
\label{2.8}
 S[\; \rho(\sgm_1)\; \rho(\sgm_2) \; ] = S[\; \rho(\sgm_1)\; ] + 
S[\; \rho(\sgm_2)\; ] \; .
\ee

\vskip 2mm
{\bf Shannon theorem} \cite{Shannon_4}. {\it The unique functional, satisfying the 
above conditions, up to a positive constant factor, has the form}:
\be
\label{2.9}
 S[\; \rho(\sgm)\; ] = - \sum_\sgm \rho(\sgm) \; \ln \; \rho(\sgm) \;  .
\ee

\vskip 2mm

Here, the natural logarithm is used, but, generally, it is not important what 
logarithm base is employed since the entropy is defined up to a constant factor. 
The information entropy coincides with the Gibbs entropy in statistical mechanics 
\cite{Gibbs_5,Gibbs_6}. For the uniform distribution, we have
\be
\label{2.10}
 S \left[\; \frac{1}{N_{tot}} \; \right] = \ln N_{tot} \; ,
\ee
which, clearly, is a monotonic increasing function of $N_{tot}$. Generally, the 
entropy is in the range
\be
\label{2.11}
 0 \leq S[\; \rho(\sgm)\; ] \leq \ln N_{tot} \;  .
\ee
It is zero, when there is just a single agent with a single characteristic, so 
that $N_{tot} =1$ and the distribution is trivial, then
\be
\label{2.12}
 S[\; 1 \; ] = 0 \qquad ( N_{tot} = 1 ) \; .
\ee
The information entropy is a measure of the system uncertainty. In other words, 
one can say that the entropy is a measure of unaccessible information.  

When a statistical system, with a distribution $\rho(\sigma)$, is initially 
characterized by a trial likelihood distribution $\rho_0(\sgm)$, then the form
\be
\label{2.13}
   I_{KL}[\; \rho(\sgm)\; ] = 
\sum_\sgm \rho(\sgm) \; \ln \; \frac{\rho(\sgm)}{\rho_0(\sgm)}
\ee
is called the Kullback-Leibler {\it information gain}, or relative entropy, or 
Kullback-Leibler divergence \cite{Kullback_7,Kullback_8}.

\subsection{Information Functional}

Information functional has to include the information gain (\ref{2.13}). In 
addition, we should not forget that the probability distribution is to be 
normalized, as in Eq. (\ref{2.3}), hence
\be
\label{2.14}
\sum_\sgm \rho(\sgm) -1 = 0 \; .
\ee
There can exist another information defining some average quantities $C_i$ by 
the condition
\be
\label{2.15}
\sum_\sgm \rho(\sgm) C_i(\sgm) - C_i = 0 \; .
\ee
Overall, the information functional takes the form
\be
\label{2.16}
 I[\; \rho(\sgm) \; ] = 
\sum_\sgm \rho(\sgm) \; \ln \; \frac{\rho(\sgm)}{\rho_0(\sgm)} +
\lbd_0 \left[ \sum_\sgm \rho(\sgm) - 1 \right] + 
\sum_i \lbd_i \left[ \sum_\sgm \rho(\sgm) C_i(\sgm) - C_i \right] \; .
\ee
  
The use of the Kullback-Leibler information gain for deriving probability 
distributions is justified by the Shore-Johnson theorem \cite{Shore_9}.

\vskip 2mm
{\bf Shore-Jonson theorem} \cite{Shore_9}. {\it There exists only one distribution 
satisfying consistency conditions, and this distribution is uniquely defined by the 
minimum of the Kullback-Leibler information gain, under given constraints}.

\vskip 2mm 

In the information functional (\ref{2.16}), the coefficients $\lbd_0$ and 
$\lbd_i$ are the Lagrange multipliers, whose variation 
$$
\frac{\prt I[\; \rho(\sgm) \; ]}{\prt\lbd_0} = 0 \; , 
\qquad
\frac{\prt I[\; \rho(\sgm) \; ]}{\prt\lbd_i} = 0 
$$
yields the normalization condition (\ref{2.14}) and the expectation conditions 
(\ref{2.15}).

\vskip 2mm
{\bf Principle of minimal information}. {\it The probability distribution of an 
equilibrium statistical system (\ref{2.4}) is defined as the minimizer of the 
information functional (\ref{2.16}), thus satisfying the conditions}:
\be
\label{2.17}
\frac{\dlt I[\; \rho(\sgm) \; ]}{\dlt\rho(\sgm)} = 0 \; , 
\qquad
\frac{\dlt^2 I[\; \rho(\sgm) \; ]}{\dlt\rho(\sgm)^2} > 0 \;  .
\ee

\vskip 2mm
The minimization conditions give
$$
\frac{\dlt I[\; \rho(\sgm) \; ]}{\dlt\rho(\sgm)} = 1 + \lbd_0 + 
\ln \; \frac{\rho(\sgm)}{\rho_0(\sgm)} + \sum_i \lbd_i C_i(\sgm) \; ,
\qquad
\frac{\dlt^2 I[\; \rho(\sgm) \; ]}{\dlt\rho(\sgm)^2} = 
\frac{1}{\rho(\sgm)} \; ,
$$
which leads to the distribution
\be
\label{2.18}
 \rho(\sgm) = \frac{\rho_0(\sgm)\exp\{-\sum_i \lbd_i C_i(\sgm)\} }
{\sum_\sgm\rho_0(\sgm)\exp\{- \sum_i \lbd_i C_i(\sgm)\} } \; .
\ee
If there is no preliminary information on the trial distribution properties, 
so that all states are equiprobable, this implies that the trial distribution 
is uniform,
$$
\rho_0(\sgm) = \frac{1}{N_{tot}} \;   .
$$
Then the sought distribution reads as
\be
\label{2.19}
\rho(\sgm) = 
\frac{1}{Z} \; \exp\left\{ - \sum_i \lbd_i C_i(\sgm)\right\} \; ,
\ee
with the normalization factor, called partition function,
\be
\label{2.20}
Z = \sum_\sgm \exp\left\{ - \sum_i \lbd_i C_i(\sgm)\right\} \; .
\ee

In statistical mechanics, this is called the Gibbs distribution 
\cite{Gibbs_5,Gibbs_6}. The principle of minimal information is equivalent to 
conditional maximization of entropy. As is stated by Jaynes, the Gibbs distribution 
can be used for any complex system whose description is based on information theory 
\cite{Jaynes_10,Jaynes_11}.

It may happen that the probability distribution $\rho(\sigma,x)$ depends on 
a parameter $x$ that has not been uniquely prescribed. In such a case, this 
parameter can be chosen such that to follow the principle of minimal 
information. To this end, substituting the distribution (\ref{2.19}) into 
the information functional (\ref{2.16}) results in
\be
\label{2.21}
 I [\; \rho(\sgm,x)\; ] = R(x) - \sum_i \lbd_i C_i \;  ,
\ee
where 
\be
\label{2.22}
R(x) = - \ln Z(x)
\ee
is {\it relative information}. Keeping in mind that the average quantities 
$C_i$ are fixed, we see that the minimization of the information functional 
with respect to the parameter $x$ is equivalent to the minimization of the 
relative information (\ref{2.22}), since 
\be
\label{2.23}
 \frac{\prt I [\; \rho(\sgm,x)\; ]}{\prt x} = 
\frac{\prt R(x)}{\prt x} = 0 \; ,
\qquad
\frac{\prt^2 I [\; \rho(\sgm,x)\; ]}{\prt x^2} = 
\frac{\prt^2 R(x)}{\prt x^2} > 0 \;  .
\ee
In that way, the probability distribution can be uniquely defined \cite{Yukalov_12}.

\subsection{Representative Ensembles}

When the probability distribution is defined by the principle of minimal 
information, this does not mean that it is useful to have little information 
on the system. Vice versa, it is necessary to include into the information 
functional all available relevant information on the system, corresponding 
to the expected-value conditions. The principle of minimal information shows 
how to obtain an optimal description of a complex system, while possessing 
the minimal information on the latter. All available important information 
on the system must be taken into account.

Thus a statistical system is described by the triple (\ref{2.4}) consisting 
of $N$ system members, society set $X$, and the probability distribution 
$\rho(\sgm)$. The pair $\{X,\rho(\sgm)\}$ is called {\it statistical ensemble}. 
Observables are represented by real functions $A(\sgm)=A^*(\sgm)$ whose averages
\be
\label{2.24}
\lgl \; A(\sgm) \; \rgl = \sum_\sgm \rho(\sgm) A(\sgm)
\ee
are the observable quantities that can be measured. The collection of all 
available observable quantities $\{\lgl A(\sgm)\rgl \}$ is a {\it statistical 
state}.  

A statistical ensemble is termed {\it representative} when it provides the 
correct description for the system statistical state, so that the theoretical 
expectation values of observable quantities accurately describe the corresponding 
measured values. For this, it is necessary to include into the information 
functional all relevant information in the form of additional constraints. 
Only then the minimization of the information functional will produce a correct 
probability distribution. 

The idea of representative ensembles goes back to Gibbs \cite{Gibbs_5,Gibbs_6}.
Their importance is emphasized by Ter Haar \cite{Ter_13,Ter_14}. The necessity 
of employing representative ensembles for obtaining reliable theoretical 
estimates has been analyzed and the explanation how the use of non-representative 
ensembles leads to incorrect results has been discussed in detail 
\cite{Yukalov_15,Yukalov_16,Yukalov_17,Yukalov_18}.

\subsection{Arrow of Time}

In general, the probability distribution $\rho(\sigma,t)$ can depend on time $t$.
An important question is: Why time is assumed to always increase? One usually 
connects this with the second law of thermodynamics according to which the 
entropy of an isolated system left to spontaneous evolution cannot decrease
\cite{Kubo_19}. However, strictly speaking, by Liouville theorem, the entropy 
of a closed system remains constant in time  (see, e.g., \cite{Yukalov_20}).
It is possible to infer that the arrow of time appears in quasi-isolated systems
due to their stochastic instability 
\cite{Yukalov_21,Yukalov_22,Yukalov_23,Yukalov_24,Yukalov_25}. Here we show that 
there is a very simple way of proving that the irreversibility of time can be 
connected with the non-decrease of information gain. First, we need to remind the
Gibbs inequality.

\vskip 2mm

{\bf Gibbs inequality}. {\it For two non-negative functions $A(\sgm)\geq 0$ and 
$B(\sgm)\geq 0$, one has}:
\be
\label{2.25}
 \sum_\sgm [ \; A(\sgm) \ln A(\sgm) - A(\sgm) \ln B(\sgm) \; ] 
\geq \sum_\sgm [\; A(\sgm) - B(\sgm) \; ] \;  .
\ee

\vskip 2mm

{\it Proof}. The proof follows from the inequality
$$
\ln \; \frac{A\sgm)}{B(\sgm)} \geq 1 \; - \; \frac{B(\sgm)}{A(\sgm)} \; .
$$

\vskip 2mm
{\it Consequence}. The information gain, caused by the transition from the 
probability $\rho_0(\sigma)$ to $\rho(\sigma)$, is semi-positive:
\be
\label{2.26}
 \sum_\sgm \rho(\sgm) \; \ln \; \frac{\rho(\sgm)}{\rho_0(\sgm)} \geq 0 \; .
\ee
More examples of useful inequalities in information theory can be found in 
\cite{Dembo_26}.

\vskip 2mm

Now let us consider the natural change of the probability distribution with 
time, starting from the initial value $\rho(\sgm,0)$ to the final $\rho(\sgm,t)$
at time $t$. Then the following statement is valid.

\vskip 2mm
{\bf Non-decrease of information gain}. {\it The information gain, due to the 
evolution of the probability distribution from the initial distribution 
$\rho(\sigma,0)$ to the final $\rho(\sigma,t)$, does not decrease}:
\be
\label{2.27}
\sum_\sgm \rho(\sgm,t) \; 
\ln \; \frac{\rho(\sgm,t)}{\rho(\sgm,0)} \geq 0 \;  .
\ee

\vskip 2mm
Thus the non-decrease of the information gain with time can be connected with 
the direction of time.

\section{Equilibrium Social Systems}

This section introduces the main notions required for modeling equilibrium 
social systems and studies some simple statistical models. In the long run, 
social systems are, strictly speaking, nonequilibrium. However, if the society 
does not experience external shocks during the period of time much longer than 
the typical interaction time between the society members, this society can be 
treated as equilibrium for that period of time. More details on the so-called 
social physics can be found in Refs. 
\cite{Weidlich_27,Parsons_28,Galam_29,Perc_30,Perc_31,Jusup_32}.

\subsection{Free Energy}

Equilibrium social systems can be treated by statistical theory being a 
particular application of the above notions of information theory and the 
principle of minimal information. Following the notation of the previous 
section, let us consider a society of $N$ members, where each member is 
associated with characteristics (\ref{2.1}). The considered society occupies 
a volume $V$.

Our aim is to study almost isolated societies, with self-organization due 
to their internal properties. Of course, no real society can be absolutely 
isolated from surrounding. The influence of surrounding is treated as stationary 
random perturbations, or stationary noise. The perturbations can be produced 
by other societies, by natural causes, such as earthquakes, floods, droughts, 
epidemics, etc. The noise is considered as being stationary, which implies 
that equilibrium situation is assumed. The influence of noise on the society 
is measured by temperature $T$ that is the measure of noise intensity. In the 
limiting cases, the absence of noise implies $T=0$, and extremely strong noise 
means $T= \infty$. One often uses the inverse temperature 
\be
\label{3.1}
 \bt \equiv \frac{1}{T}  
\ee
that can be interpreted as the measure of society isolation from random noise. 
Respectively, $\bt=0$ means no isolation and extreme noise, while $\bt=\infty$ 
implies the complete isolation and no noise. 

An equilibrium system is conveniently characterized by a functional $H(\sgm)$ 
termed Hamiltonian, or harm. The expected value of the Hamiltonian is the 
society energy
\be
\label{3.2}
 E = \sum_\sgm \rho(\sgm) H(\sgm) \equiv \lgl \; H(\sgm) \; \rgl \; ,
\ee
which is also called society cost. In some economic and financial applications 
it is termed disagreement or dissatisfaction, since the higher energy assumes 
the more excited society.

The energy of noise is $TS$, with $S$ being the information entropy. The free 
energy is the part of the society energy, due to the society itself, without the 
noise energy:
\be
\label{3.3}
F = E - TS \; .
\ee
In the limiting case of zero temperature, when there is no noise, the free 
energy coincides with the society energy,
\be
\label{3.4}
 F = E \qquad ( T = 0 ) \;  .
\ee

The probability distribution over the society characteristics is defined by the 
principle of minimal information, which in the present notation gives
\be
\label{3.5}
 \rho(\sgm) = \frac{1}{Z} \; e^{-\bt H(\sgm)} \;  ,
\ee
with the partition function
\be
\label{3.6}
 Z = \sum_\sgm e^{-\bt H(\sgm)} \;  .
\ee

Substituting distribution (\ref{3.5}) into entropy (\ref{2.9}) leads to
\be
\label{3.7}
 S = \bt E + \ln Z \; .
\ee
Comparing (\ref{3.7}) with (\ref{3.3}) results in the expression
\be
\label{3.8}
 F = - T \; \ln Z \; .
\ee
It is easy to check that the entropy $S$ can be represented as
\be
\label{3.9}
  S = -\; \frac{\prt F}{\prt T} \; .
\ee

Society pressure is defined as
\be
\label{3.10}
 P = -\; \frac{\prt F}{\prt V} \;  .
\ee
This relation implies that the region occupied by a society can be changed 
because of pressure. Population density is
\be
\label{3.11}
 \rho \equiv \frac{N}{V} \;   .
\ee

Combining (\ref{2.22}) and (\ref{3.8}) gives the equality
\be
\label{3.12}
F(x) =  - T \ln Z(x) = T R(x) \; .
\ee
Hence the information functional (\ref{2.21}) and the free energy (\ref{3.12}) 
are connected:
\be
\label{3.13}
 I[\;\rho(\sgm,x)\;] = \bt F(x) - \sum_i \lbd_i C_i \; .
\ee
Since the values $C_i$ are fixed, we have
\be
\label{3.14} 
 \frac{\prt I[\;\rho(\sgm,x)\;]}{\prt x} = 
\frac{1}{T} \; \frac{\prt F(x)}{\prt x} \;  .
\ee
Therefore the minimization of the information functional over a parameter $x$ 
is equivalent to the minimization of free energy over this parameter:
\be
\label{3.15}
\min_x I[\;\rho(\sgm,x)\;] \longleftrightarrow \min_x F(x) \; .
\ee

\subsection{Society Stability}

Considering society models, it is important to make it sure that the society 
is stable. A society is stable if small variations of parameters do not drive 
it far from its initial state. Between several admissible statistical states, 
the system chooses that which provides the minimal free energy. The system is 
stable with respect to the variation of parameters if its free energy is 
minimal, so that
\be
\label{3.16}
 \dlt F = 0 \; , \qquad \dlt^2 F > 0 \; .
\ee

The variation of the intensity of noise, that is of temperature, under fixed 
volume, is characterized by specific heat
\be
\label{3.17}
C_V = \frac{1}{N} \left( \frac{\prt E}{\prt T}\right)_V = - \; 
\frac{T}{N} \; \left( \frac{\prt^2F}{\prt T^2} \right)_V \;  .
\ee
With the free energy (\ref{3.8}), this takes the form 
\be
\label{3.18}
C_V = \frac{{\rm var}(H)}{NT^2} \; ,
\ee
where the variation of the Hamiltonian $H = H(\sigma)$ means
\be
\label{3.19}
 {\rm var}(H) \equiv \lgl \; H^2 \;\rgl - \lgl \; H\;\rgl^2 \;  .
\ee
The latter is non-negative, hence the specific heat has to be non-negative.

The variation of volume, under fixed temperature, is characterized by the 
compressibility
\be
\label{3.20}
 \varkappa_T = -\; \frac{1}{V} \left( \frac{\prt V}{\prt P} \right)_{TN}
=  \frac{1}{V} \left( \frac{\prt^2 F}{\prt V^2} \right)_{TN}^{-1} \; .
\ee

The condition of stability requires that the specific heat, as well as 
compressibility, be positive and finite,
\be
\label{3.21}
0 \leq C_V < \infty \; , \qquad   0 \leq \varkappa_T < \infty \; .
\ee
In this way, the principle of minimal information agrees with the minimal 
free energy, which, in turn, assumes the society stability.

As an example of society instabilities, it is possible to mention 
disintegration of a country as a result of a war or because of external 
economic pressure. Another example is bankruptcy of a firm caused by changed 
financial conditions.

\subsection{Practical Approaches}

In order to accomplish quantitative investigations of society properties, two 
different approaches are employed: network approach and typical-agent approach.

The network approach, also called multi-agent modeling, is based on the 
following assumptions:

\vskip 2mm
(i) The considered society consists of agents, or nodes, that are fixed at the 
lattice sites of a spatial (usually two-dimensional) lattice. 

\vskip 2mm
(ii) The agents interact with each other when they are close to each other, 
usually the nearest-neighbor interactions are considered.

\vskip 2mm
(iii) Because of rather complicated calculations, as a rule, one has to resort 
to numerical modeling with computers.

\vskip 2mm

There exist many examples of networks, such as electric-current networks, models 
of magnetic and ferroelectric materials, neuron networks in brains, computer 
networks, and so on. 

The results of the network approach depend on lattice dimensionality 
modeling the society (whether one-, or two-, or three-dimensional lattices 
are considered), lattice geometry (cubic, triangular, or another structure), 
and on the interaction type and range (long-range, short-range, mid-range). 
The network approach is appropriate for small or simply structured systems, 
with agents that can be treated as fixed at spatial points. 

However complex societies, such as human and biological societies, are not 
composed of agents tied to a spatial lattice, usually they are not attached 
to any fixed spatial locations or sites. The agent interactions are not of 
nearest-neighbor type and can be independent of distance between the agents. 
Interactions in a society can be either direct, when interacting with fixed 
neighbors, or can involve changing neighbors. Nowadays the interactions not 
depending on the distance are widespread, such as through phone, Skype, 
WhatsApp, Telegram, and like that. There exist indirect interactions through 
letters or e-mails, by reading news papers and books, by listening to radio 
or watching television. The majority of interactions are long-range, but not 
solely with nearest neighbors. Other biological societies are also interacting 
at large distances by means of their voices and smells.  

Summarizing, complex societies, like human or other biological societies, are 
formed by agents that are not fixed at spatial locations and can interact at 
long distance. This kind of societies is better described by the typical-agent 
approach. In this approach, one reduces the problem to the consideration of 
the behavior of typical agents representing a kind of an average member of 
the society. For example, the typical interaction of agents, having the 
characteristics $\sgm_i$ and $\sgm_j$, and described by the term $\sgm_i\sgm_j$, 
is transformed to the expression
\be
\label{3.22}
 \sgm_i \sgm_j = \sgm_i \lgl \; \sgm_j \;\rgl + 
\lgl \; \sgm_i \;\rgl \sgm_j - 
\lgl \; \sgm_i \;\rgl \lgl \; \sgm_j \;\rgl\; ,
\ee
in which $i \neq j$. The average characteristic is
\be
\label{3.23}
 \lgl \; \sgm_j \;\rgl = \sum_\sgm \rho(\sgm) \sgm_j \equiv s \; .
\ee
Thus the representation (\ref{3.22}) becomes
\be
\label{3.24}
 \sgm_i \sgm_j = s \sgm_i + s \sgm_j - s^2  
\ee
and the average interaction reads as
\be
\label{3.25}
\lgl \; \sgm_i \sgm_j \;\rgl = 
\lgl \; \sgm_i \;\rgl  \lgl \; \sgm_j \;\rgl = s^2 \; .
\ee

Then the description of the interaction is reduced to the consideration of 
typical agents subject to the average influence of other typical agents. For 
complex social systems, the typical-agent approach is not merely simpler but 
is more correct.

\subsection{Society Transitions}

Usually, one is interested not in the characteristics of single agents but 
in the general behavior of a society. For this purpose, one considers the mean 
arithmetic characteristic
\be
\label{3.26} 
s(\sgm) = \frac{1}{N} \sum_{j=1}^N \sgm_j \;  .
\ee
The observable quantity is the average characteristic
\be
\label{3.27}
 s \equiv \lgl \; s(\sgm) \;\rgl = 
\frac{1}{N} \sum_{j=1}^N \lgl \; \sgm_j \;\rgl \; .
\ee
 
The general behavior of a society can be associated with its average 
characteristics. When the property of the characteristic qualitatively changes, 
one speaks of a {\it social phase transition}. For example, it may happen that 
under such a social transition the average characteristic varies between zero 
and nonzero values. Conditionally, one can name the social state with $|s| > 0$ 
an ordered state, while the state, where $s = 0$, a disordered state. In that 
case, the average characteristic (\ref{3.27}) is termed {\it order parameter}. 
There can occur the following types of transitions when a system parameter, for 
instance temperature, varies.

\vskip 3mm
{\it First-order transition}. This type of a transition happens when the 
order parameter at some point $T_0$ changes between nonzero and zero by a 
discontinuous jump:
\be
\label{3.28}
s(T_0-0) > 0 \; , \qquad s(T_0+0) = 0 \; , 
\qquad
 s(T_0-0) \neq s(T_0+0) \;  .
\ee
Discontinuous social transitions can be associated with revolutions.

\vskip 2mm
{\it Second-order transition}. Then the order parameter at a critical point 
$T_c$ changes between nonzero and zero continuously:
\be
\label{3.29}
s(T_c - 0) > 0 \; , \qquad s(T_c + 0) = 0 \; , 
\qquad
 s(T_c - 0) = s(T_c + 0) \; .
\ee
Continuous transitions describe evolutions.

\vskip 2mm
{\it Gradual crossover}. The order parameter does not become zero at a finite 
point, but at some crossover point $T_c$ it strongly diminishes and tends to 
zero only in the limit of large $T$:
$$
s(T_c - 0) = s(T_c + 0) > 0 \; , 
$$
\be
\label{3.30}
\qquad s(T) \ra 0 \qquad ( T\gg T_c) \;  .
\ee
This transition corresponds to a smooth evolution. 

\vskip 2mm
 
There can occur more unusual situations, when a society is not completely 
equilibrium \cite{Yukalov_33}, but for the description of equilibrium societies, 
the above three types of social transitions are sufficient.

\subsection{Yes-No Model}

In the present section, we consider a simple model, that is well known in 
statistical physics where it is called Ising model 
\cite{Huang_44,Kubo_45,Isihara_46,Kadanoff_2000}. This model is also often used 
in different applications to financial and economic problems \cite{Sornette_80}. 
We need this model in order to illustrate in action the notions introduced in the 
previous sections, to exemplify the terminology associated with social systems, 
and for having the ground for studying more complicated models in the following 
sections.  

Suppose each agent of a society can have just two features, which are opinions 
that can be termed "yes" and "no". Generally, this can be any decision with 
two alternatives. For instance, this can be voting for or against a candidate 
in elections, supporting or rejecting a suggestion in a referendum, buy or sell 
stocks in a market, etc. These two alternatives can be represented by the binary 
variable taking two possible values, e.g.:
\begin{eqnarray}
\label{3.31}
\sgm_j = \left\{ \begin{array}{ll}
-1 \; , ~ & ~ no \\
+1 \; , ~ & ~ yes
\end{array}
\right. \; .
\end{eqnarray}

The interaction, or mutual influence between two members of the society writes 
as
\be
\label{3.32}
 H_{ij} = - J_{ij} \sgm_i \sgm_j \; ,
\ee
with the value $J_{ij} = J_{ji}$ being the intensity of the interaction. The 
case of agreement (collaboration) or disagreement (competition) of the members, 
respectively, corresponds to the values
$$
J_{ij} > 0 \qquad (agreement , ~ collaboration) \; ,
$$
\be
\label{3.33}
J_{ij} < 0  \qquad (disagreement , ~ competition) \; .
\ee
The terminology comes from the fact that, under agreement, when $J_{ij}>0$, 
the interaction energy is minimal for coinciding $\sgm_i$ and $\sgm_j$, while 
in the case of disagreement, when $J_{ij}<0$, the interaction energy is minimal 
for opposite $\sgm_i$ and $\sgm_j$. When there is mutual agreement and both 
agents vote in the same way, either both "yes" or both "no", the interaction 
energy is lower than when the agents would vote differently. 

The Hamiltonian of the system is
\be
\label{3.34}
H = - \; \frac{1}{2} \sum_{i\neq j}^N J_{ij} \sgm_i \sgm_j \;  .
\ee
In physics, this is called the Ising model (see the history of the model in 
\cite{Brush_34}).

In the typical-agent approach, the Hamiltonian reads as
\be
\label{3.35}
 H = - Js \sum_{j=1}^N \sgm_j + \frac{1}{2} \; J s^2 N  \; ,
\ee
where
\be
\label{3.36}
 J \equiv \frac{1}{N} \sum_{i\neq j}^N J_{ij} = 
\sum_{i(\neq j)}^N J_{ij} \;  .
\ee

Using the equalities
$$
\exp\left( - \bt J s \sum_{j=1}^N \sgm_j \right) = 
\prod_{j=1}^N \exp(-\bt J s \sgm_j) \; ,
$$
$$
\sum_\sgm \prod_{j=1}^N \exp(-\bt J s \sgm_j) =
[ \; \exp(\bt J s) + \exp(- \bt J s ) \; ]^N \; ,
$$
we get the probability distribution
\be
\label{3.37}
 \rho(\sgm) = \prod_{j=1}^N \rho(\sgm_j) \; ,
\ee
in which
\be
\label{3.38}
 \rho(\sgm_j) = 
\frac{\exp(\bt J s\sgm_j)}{\exp(\bt J s) + \exp(-\bt J s)} \;  .
\ee

For the order parameter (\ref{3.27}), we obtain the equation
\be
\label{3.39}
 s = \tanh(\bt J s)  \; .
\ee

When there is mutual disagreement between the agents, so that $J<0$, the sole 
solution for the order parameter is $s=0$, which means a disordered state,
\be
\label{3.40}
 s\equiv 0 \qquad ( J < 0 ) \; .
\ee

However, when the members of the society are in mutual agreement, so that $J>0$, 
then there are two solutions of Eq. (\ref{3.38}). One solution is $s=0$, but the 
other solution is nonzero for $T<T_c=J$. In order to choose the stable solution, 
we need to find out which of them minimizes the free energy. 

It is convenient to work with the dimensionless reduced free energy
\be
\label{3.41}
 \overline F(s) \equiv \frac{F}{NJ} \qquad ( J > 0 )  
\ee
and to measure temperature in units of $J$, keeping in mind a positive $J$. Then 
we have
\be
\label{3.42}
 \overline F(s)  = \frac{1}{2} \; s^2 - T \ln [\; 2 \cosh(\bt s) \; ] \; .
\ee
The condition of the free-energy extremum is
\be
\label{3.43}
 \frac{\prt \overline F(s)}{\prt s} = s - \tanh(\bt s) = 0 \;  ,
\ee
which gives the order-parameter equation (\ref{3.39}). The condition of the 
free-energy minimum
\be
\label{3.44}
 \frac{\prt^2 \overline F(s)}{\prt s^2} = \tanh^2(\bt s) > 0 
\ee
holds true for nonzero $s$. It is also not difficult to check that 
\be
\label{3.45}
   \overline F(s) <  \overline F(0) \qquad ( T < T_c) \; ,
\ee
where $\overline{F}(0) = -T \ln 2$. Hence, for temperatures lower than the 
critical temperature $T_c = J >0$, the stable state corresponds to the nonzero 
value of the order parameter $s$. 

At zero temperature $(T = 0)$, with collaborating agents $(J>0)$, the society 
is completely ordered:
\be
\label{3.46}
 s = \pm 1 \qquad ( T = 0 \; , ~ J > 0 ) \;  ,
\ee
although with an unspecified decision, either all deciding "yes" or all choosing 
the option "no".

It is easy to notice that for any $T$ below $T_c$ there are two solutions, 
positive and negative, both corresponding to the same free energy 
$\overline{F}(s)=\overline{F}(-s)$ that does not depend on the sign of the 
order parameter. This means that with equal probability the society can vote 
"yes" as well as "no". In that sense, the situation is degenerate.

\subsection{Enforced Ordering}

The yes-no degeneracy can be lifted by imposing an ordering force acting on the 
members of the society. The ordering force, or regulation force, acting on the 
members, represents different regulations, such as governmental rules and laws, 
as well as the society traditions and habits. The society Hamiltonian, including 
the ordering force, takes the form
\be
\label{3.47}
 H = - \; \frac{1}{2} \sum_{i\neq j}^N J_{ij} \sgm_i \sgm_j - 
\sum_{j=1}^N B_j \sgm_j \;  .
\ee
I what follows, we assume that the regulation is uniform, that is that the same 
regulations are applied to all members of the society. This translates into the 
condition that the ordering force is the same for all agents, such that
\be
\label{3.48}
 B_j = B_0 \;  .
\ee
In other words, the laws are the same for everyone. 

Resorting to the typical-agent approach yields the Hamiltonian
\be
\label{3.49}
 H = - J ( s + h ) \sum_{j=1}^N \sgm_j + \frac{1}{2} \; J s^2 N \; ,
\ee
with the dimensionless force
\be
\label{3.50}
 h \equiv \frac{B_0}{J} \qquad ( J > 0 )  
\ee
and the order parameter 
\be
\label{3.51}
 s \equiv \lgl \; \sgm_j \; \rgl \;  .
\ee
We keep in mind the case of mutual agreement, where $J>0$.

In the typical-agent approach, the reduced free energy becomes
\be
\label{3.52}
 \overline F(s) = \frac{1}{2} \; s^2 - 
T \ln \left[ \; 2\cosh\left( \frac{s+h}{T}\right) \; \right] \; ,
\ee
where temperature is measured in units of $J$. The order parameter (\ref{3.51}) 
is given by the equation
\be
\label{3.53}
s = \tanh \left( \frac{s+h}{T}\right) \;   .
\ee
The order parameter can also be defined as the derivative
\be
\label{3.54}
 s = - \; \frac{\prt \overline F(s)}{\prt h} \;  .
\ee
 
How the society responds to the imposed regulations is described by the 
susceptibility
\be
\label{3.55}
\chi = \frac{\prt s}{\prt h} = - \; 
\frac{\prt^2\overline F(s)}{\prt h^2}   
\ee
that can be represented as
\be
\label{3.56}
 \chi = \frac{1}{T} \; {\rm var}(s(\sgm) ) \; ,
\ee
with the variance
$$
 {\rm var}(s(\sgm) ) = \lgl \; s^2(\sgm) \; \rgl - 
\lgl \; s(\sgm) \; \rgl^2 \; .
$$
Thus the susceptibility is positive, which means that the imposed regulations 
increase the order. Explicitly, we find
\be
\label{3.57}
 \chi = \frac{1-s^2}{T+s^2-1} \; .
\ee

From the order-parameter equation (\ref{3.53}), it is seen that the sign of 
the order parameter is prescribed by the sign of the ordering force, so that 
$s>0$ for $h>0$ and $s<0$ for $h<0$. By choosing the sign of $h$ it is possible 
to enforce either the ordering "yes" or the ordering "no". For concreteness, 
we consider $h>0$. If the ordering force is extremely strong, then
\be
\label{3.58}
 s \simeq 1 \qquad ( h \ra \infty) \; .
\ee

In that way, the imposed ordering force increases the order in the society and 
makes it more stable. For illustration, let us consider the case of no noise, 
hence $T=0$. Then the reduced free energy coincides with the reduced energy,
\be
\label{3.59}
 \overline E \equiv \frac{E}{JN} = - \; \frac{1}{2} \; - \; h
\qquad ( T = 0 ) \; .
\ee
As is seen, the imposed force diminishes the society energy, which assumes that 
the society should be more stable.

\subsection{Command Economy}

From the previous section, it looks that the stricter the regulations, the 
more the society ordered. That is, the larger the force $h$, the larger the 
order parameter $s$. The seeming conclusion could be that it is profitable to 
make the regulations as stringent as possible. Is this so? Let us consider as 
an example an economic society. The most strictly regulated type of economic 
organization is command economy or centrally planned economy. Whether such an 
overregulated economics is the most efficient one? 

The basic points of command economy are: 

\begin{enumerate}[label=(\roman*)]
\item
A centralized government owns most means of production and businesses. 

\item
Government controls production levels and distribution quotas.

\item
Government controls all prices and salaries.
\end{enumerate}

The proclaimed advantages of command economy assume that regulatory decisions 
are made for the benefit of the whole society, there is no large economic 
inequality, and the economy is claimed to be more stable. However in reality 
the proclaimed catchwords confront a huge amount of problems: 

\begin{enumerate}
\item
It is not always well defined what are the benefits of the society. 
Governmental decision makers often take decisions in their own favor, but 
not in favor of the society, announcing their egoistic goals as the society 
objective needs.

\item
It is practically impossible to formulate correct plans for all goods 
for long future. Constant shortages of necessary goods and surpluses of 
unnecessary goods are a rule. Economy is in a permanent crisis.

\item
Since the conditions for long-term future production cannot be exactly predicted,
the plans are never accomplished. The necessity of correcting the plans, with 
additional spending for the plan corrections. 

\item
Since everything is planned in advance, it is practically impossible to 
introduce innovations that have not been planned. As a result, technological 
retardation is imminent.

\item
What kind of science to develop is also planned. Some sciences are 
mistakenly announced to be unnecessary or wrong. Examples are cybernetics 
and genetics in the Soviet Union. This results in an irreparable harm to 
economy.

\item
It is impossible to absolutely farely distribute wealth. Those who 
distribute always take more for themselves. Consequently, people are 
dissatisfied. Bureaucratic corruption is flourishing leading to enormous 
economic losses.

\item
Because wealth is rigidly distributed by the government, there is no 
any reason to work hard. Labor becomes inefficient, with low productivity.

\item
The necessity of having huge planning institutions consumes a large 
amount of economic means. Ineffective planning reduces the planners to 
the state of parasites, merely wasting resources.

\item
The necessity of having a large number of controllers for implementing 
economic plans and controlling their accomplishment also makes such 
controllers as parasites.

\item
Suppression of economic freedom, ascribed to the needs of the state, 
but usually needed for protecting the privileges of the country rulers, 
kills the motivation of people to work well.

\item
Because of the total frustration of suppressed citizens, there is 
the necessity of having excessively large police and regulating services 
supervising the society.

\item
To support the order and punish those who are against, or hesitate, 
or even could be potentially dangerous, the government organizes massive 
suppression of people, which results in large economic losses.

\item
For realizing unreasonable plans, government practices massive arrests 
of innocent people for creating slave labor. But slavery is not economically 
efficient.

\item
To distract the discontent of population from economic failures, the
necessity arises of inventing enemies, which requires a large army consuming 
substantial amount of the country's wealth.

\item
In order to persuade people that everything is all right, it is necessary 
to organize propaganda through mass media, which results in an ineffective 
spending of resources. 
\end{enumerate}

More details on the societies with command economy can be found in literature 
\cite{Hayek_35}. Because of so many factors making the economy of an 
overregulated society ineffective, it does not look feasible that enforcing 
regulations unlimitedly could result in an ideally ordered society. It looks 
that the model of a regulated society in the previous section does not take 
into account some factors preventing the indefinite increase of order by 
overregulating the society.

\subsection{Regulation Cost}

The problem with regulation is that it requires the use of society resources. 
Regulations are costly. The more strict regulations, the more resources 
required. All regulations, imposed on a society, are produced by a part of 
the same society, that is by the society itself. Some members of the society 
act on other members. The regulation cost is the cost the society has to pay 
in order to introduce the desired regulations. The regulation cost, keeping 
in mind the said above, can be modeled by the expression
\be
\label{3.60}
H_{reg} = \frac{1}{2} \sum_{i\neq j}^N A_{ij} B_i B_j \sgm_i \sgm_j \; ,
\ee
where $A_{ij} = A_{ji} > 0$. The coefficients $A_{ij}$ describe the regulation 
efficiency. By the order of magnitude, they are proportional to the interaction 
between the agents $J_{ij}$, since regulations need to overcome mutual 
interactions in order to impose restrictions. Note that $J_{ij}$ and $A_{ij}$ 
are different types of interactions, one is a direct interaction, not related 
to the process of ordering, the other is an interaction inducing the ordering 
in the presence of an additional force. Different types of interactions, in general, 
are different.

Thus the total Hamiltonian becomes
\be
\label{3.61}
H =  -\; \frac{1}{2} \sum_{i\neq j}^N J_{ij} \sgm_i \sgm_j -
\sum_{i=1}^N B_i \sgm_i + 
\frac{1}{2} \sum_{i\neq j}^N A_{ij} B_i B_j \sgm_i \sgm_j \;   .
\ee
We again assume that the force acting on the society members is uniform in 
the sense of equality (\ref{3.48}). 
   
Now the order parameter (\ref{3.51}) differs from the expression 
\be
\label{3.62}
M = - \; \frac{\prt\overline F}{\prt h} = s - \; 
\frac{1}{N} \sum_{i\neq j}^N \al_{ij} h \lgl \; \sgm_i \sgm_j \; \rgl \; ,
\ee
with the notation
\be
\label{3.63}
 \al_{ij} \equiv A_{ij} J  \; .
\ee
This expression (\ref{3.62}) plays the role of the total order parameter, 
contrary to $s$ that in the present case is a partial order parameter.

In the typical-agent approach, Hamiltonian (\ref{3.61}) reads as
\be
\label{3.64}
H = - (J s + B_0 - AB_0^2 s ) \; \sum_{i=1}^N \sgm_i +
\frac{1}{2} \; \left( J - A B_0^2 \right) s^2 N  \;  ,
\ee
where
\be
\label{3.65}
 A \equiv \frac{1}{N} \sum_{i\neq j}^N A_{ij} \;  .
\ee

The reduced free energy takes the form
\be
\label{3.66}
\overline F = \frac{1}{2} \; \left( 1 - \al \ h^2 \right) s^2 -
T\ln \left\{ 2 \cosh\left[ \frac{(1-\al h^2)s + h}{T} \right] \right\} \; ,
\ee
in which
\be
\label{3.67}
\al \equiv \frac{1}{N} \sum_{i\neq j}^N \al_{ij} = A J \; .
\ee
The parameters $\alpha$ and $h$ are assumed to be positive. 

The order parameter $s = \langle \sigma_j \rangle$ can be obtained from the 
condition
\be
\label{3.68}
\frac{\overline F}{\prt s} = 0
\ee
leading to the equation
\be
\label{3.69}
 s = \tanh\left[ \frac{(1-\al h^2)s + h}{T} \right] \; .
\ee
However, in the presence in the Hamiltonian of a quadratic with respect to 
the ordering force term, the quantity $s$ does not define the total order that 
is given by the form (\ref{3.62}), which yields the total order parameter
\be
\label{3.70}
 M = - \frac{\prt\overline F}{\prt h} = s - \al h s^2 \; .
\ee

The susceptibility is defined by the derivative
\be
\label{3.71}
\chi \equiv \frac{\prt M}{\prt h}  = - \; \frac{\prt^2 M}{\prt h^2}
\ee
resulting in the expression
\be
\label{3.72}
\chi = \frac{(1-s^2)(1-2\al h s)^2}{T+(1-s^2)(\al h^2-1)} \; ,
\ee
which differs from the derivative
\be
\label{3.73}
\frac{\prt s}{\prt h} = 
\frac{(1-s^2)(1-2\al h s)}{T+(1-s^2)(\al h^2-1)} \; .
\ee

The behavior of the social system at small and large regulating force shows 
how the order parameters and free energy vary. The analysis of the asymptotic 
behavior demonstrates the relation between the system stability and its order. 
This relation can be rather complicated, depending on the system parameters. 

At small ordering force and weak noise, when $0<T<1$, the order parameter $s$ 
behaves as
\be
\label{3.74}
 s \simeq s_0 + \frac{1-s_0^2}{s_0^2+T-1} \; h \qquad
( 0 < T < 1 \; , ~ h \ra 0 ) \; ,
\ee
where $s_0$ is the solution to the equation
\be
\label{3.75}
 s_0 = \tanh \left( \frac{s_0}{T} \right) \; .
\ee
The total order parameter $M$ at small ordering force and weak noise behaves as
\be
\label{3.76}
M \simeq s_0 + \left( \frac{1-s_0^2}{s_0^2+T-1} \; - \; \al s_0^2\right) h
\qquad
 ( 0 < T < 1 \; , ~ h \ra 0 ) \;  ,
\ee
which shows that the order parameters can either increase or decrease with 
rising $h$, depending on the system parameters. 

The free energy at small ordering force and weak noise is
\be
\label{3.77}
\overline F \simeq \frac{s_0^2}{2} \; - \; 
T \ln \left[ 2 \cosh\left( \frac{s_0}{T}\right) \right] - s_0 h 
\qquad 
( 0 < T < 1 \; , ~ h \ra 0 ) \;  .
\ee

For $T>1$ and weak $h$, the order parameters are 
$$
s \simeq \frac{1}{T-1}\; h \; - \; \frac{3\al(T-1)^2+T}{3(T-1)^4}\; h^3 \; , 
$$
\be
\label{3.78}
 M \simeq \frac{1}{T-1}\; h \; - \left[ \frac{3\al(T-1)^2+T}{3(T-1)^4} +
\frac{\al}{(T-1)^2} \right] h^3
\qquad
( T > 1 \; , ~ h \ra 0 ) \;  .
\ee
The free energy reads as
\be
\label{3.79}
 \overline F \simeq - T \ln 2 - \; \frac{1}{2(T-1)}\; h^2 
\qquad
 ( T > 1 \; , ~ h \ra 0 ) \; .
\ee

When the regulating force is strong, then, at any temperature, we have the order 
parameters
\be
\label{3.80}
s \simeq \frac{1}{\al h} \; - \; \frac{T-1}{\al^2 h^3} \; , \qquad
M \simeq \frac{2(T-1)}{\al^2 h^3} \qquad
( h \ra \infty)   
\ee
and the free energy
\be
\label{3.81}
\overline F \simeq - T \ln 2 \; - \; \frac{1}{2 \al} \; + \;
\frac{T-1}{2\al^2h^2}  \qquad 
( h \ra \infty) \; .
\ee

In the absence of noise, that is when $T=0$, the free energy equals the energy. 
Considering the reduced energy
\be
\label{3.82}
\overline E \equiv \frac{\lgl H \rgl}{NJ} = -\; \frac{1}{2} \; s^2
- hs + \frac{1}{2} \; \al h^2 s^2 \; ,
\ee
at zero $T$, we obtain
\be
\label{3.83}
 \overline E = - \; \frac{1}{2} \; - \; h \; + \; \frac{1}{2} \; \al h^2
\qquad ( T = 0 ) \; .
\ee
As is seen, the energy (\ref{3.83}), with switching on regulations, first 
decreases, making the society more stable,
\be
\label{3.84}
 \overline E \simeq -\; \frac{1}{2} \; - \; h \qquad
( T = 0 \; , ~ h \ra 0 ) \; .
\ee
Then it reaches the minimum
\be
\label{3.85}
 \min_h  \overline E = -\; \frac{1}{2} \left( 1 + \frac{1}{\al} \right)
\qquad
\left( h = \frac{1}{\al} \right) \; ,
\ee
after which it increases making the society less and less stable,
\be
\label{3.86}
 \overline E \simeq \frac{1}{2} \; \al h^2 \qquad 
( T = 0 \; , ~ h \ra \infty) \;  .
\ee
Thus, in the absence of noise, the optimal regulation, making the society 
the most stable, corresponds to $h=1/\alpha$, when the energy (\ref{3.85}) 
is minimal. This implies that some amount of regulations is useful, stabilizing 
the society. However, the society should not be overregulated. Overregulation 
can make the society unstable.

In the presence of noise, the situation is more complicated, being dependent 
on the noise strength and the value of the regulating force. The peculiarity 
of the system behavior is illustrated in Figs. 1 and 2. Figure 1 shows the 
behavior of the free energy $\overline{F}$ as a function of the regulation force 
$h$ for different temperatures (noise intensity), when $T<1$ in Fig. 1a and when 
$T>1$ in Fig. 1b. And Figure 2 shows the order parameter $s$ as a function of 
the regulation force $h$ for different temperatures, low $(T<1)$ and high 
$(T>1)$. As we see, at low temperature, the free energy, first, decreases an 
then increases, similarly to the behavior of energy (\ref{3.82}). The order 
parameter, first, increases but later decreases. That is, there exists an 
optimal regulation strength when the society is the most stable and at the 
same time the most ordered. At the presence of strong noise, the free energy 
decreases with $h$, however the order parameter, first increases, but then 
decreases. So, the regulating force stabilizes the society, however improves 
the order only at a limited value of $h$, while at a very strong force the 
order diminishes.

\begin{figure}[ht]
\centerline{
\hbox{ \includegraphics[width=7cm]{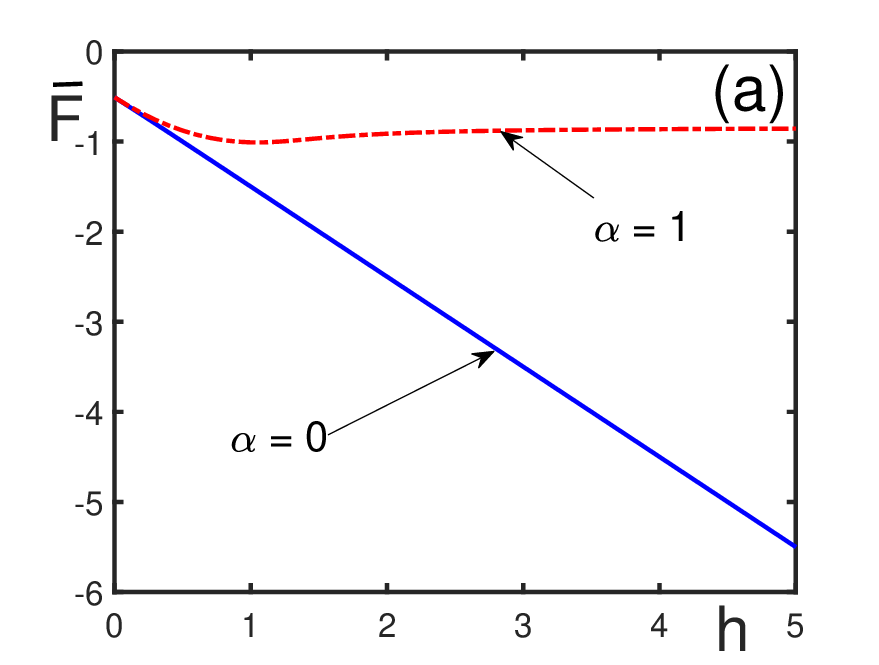} \hspace{0.5cm}
\includegraphics[width=7cm]{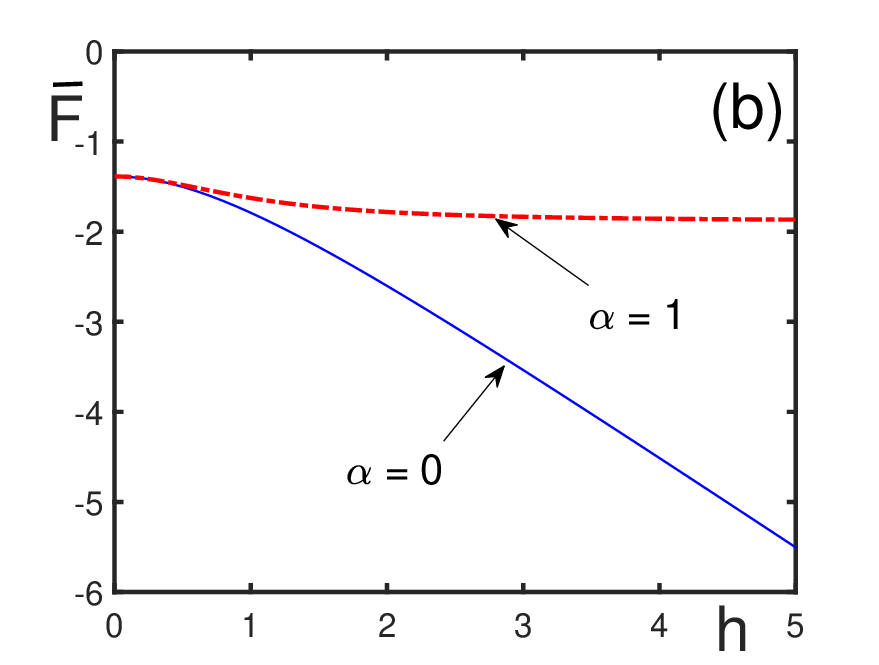}  } }
\caption{\small
Yes-no model with regulation cost. Behavior of the free energy $F$ 
as a function of the regulation force $h$ for different temperatures 
(noise intensity), when $T < 1$ in Fig. 1a and when $T > 1$ in Fig. 1b.
}
\label{fig:Fig.1}
\end{figure}

\begin{figure}[ht]
\centerline{
\hbox{ \includegraphics[width=7cm]{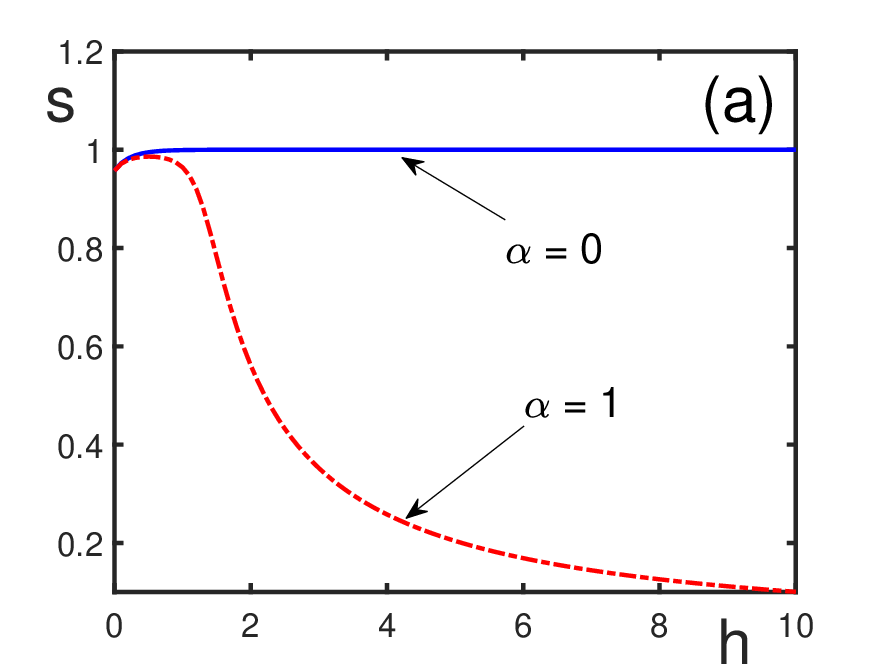} \hspace{0.5cm}
\includegraphics[width=7cm]{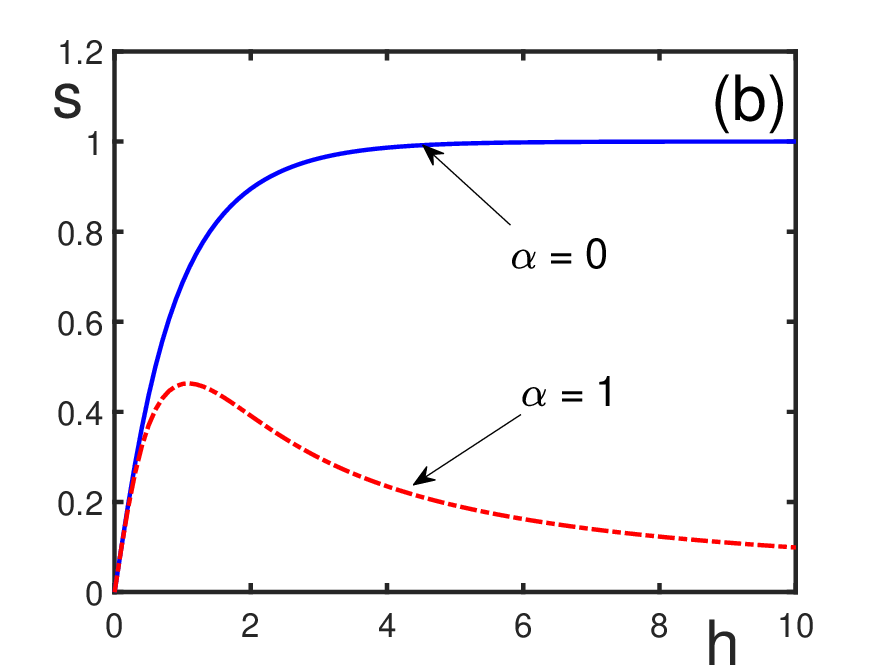}  } }
\caption{\small
Yes-no model with regulation cost. Order parameter $s$ as a function 
of the regulation force $h$ for different temperatures, low $(T < 1)$ and 
high $(T > 1)$. 
}
\label{fig:Fig.2}
\end{figure}

\subsection{Fluctuating Groups}

It is the usual situation when the society members separate into several 
groups with different properties, so that the number of members in a group 
is not constant but varies with time. For example, these could be the groups 
of opponents to the government, workers on strike, opposition organizations, 
and like that. Different groups of agents can be described by different order 
parameters. The groups can be localized in some spatial parts of the society 
or may move through the whole society space. The groups are not permanent in 
their agent numbers and do not necessarily exist forever, but they can arise 
and vanish. In that sense, the groups are fluctuating: They can appear, change,
and then disappear.

Let the number of agents in a fluctuating group be $N_f$ and the characteristic 
time during which they do not essentially change be $t_f$ . Fluctuating groups 
are mesoscopic, at least in one of the following senses.

\vskip 2mm

{\it Mesoscopic in size}:
\be
\label{3.87}
 1 \ll N_f \ll N \;  .
\ee
The number $N_f$ is much larger than one for the group order parameter to be 
defined. And $N_f$ is much smaller than the total number of agents N in the 
society to be classified as a group inside that society.

\vskip 3mm
{\it Mesoscopic in time}:
\be
\label{3.88}
t_{int} \ll t_f \ll t_{exp} \;  ,
\ee
where $t_{int}$ is the characteristic interaction time between the agents 
and $t_{exp}$ is the observation (experiment) time during which the society 
is studied. The time $t_f$ has to be much larger than the interaction time 
in order that the groups as such could be formed. And $t_f$ is much smaller 
than the observation time for the groups to be classified as fluctuating. 
The general method of describing this kind of a society containing fluctuating 
groups is presented in this section.

Let us consider a snapshot of a society picture, where the spatial location 
of a $j$-th agent of a society is denoted by a vector ${\bf a}_j$. The overall 
society containing all its members is given by the collection
\be
\label{3.89}
\mathbb{G} = \left\{ \ba_j: ~ j = 1,2, \ldots, N \right\} \; .
\ee
The society consists of several groups, each being characterized by its 
specific feature. The features are enumerated by the index $f=1,2,\ldots$. 
A group with an $f$-th feature is
\be
\label{3.90}
 \mathbb{G}_f = 
\left\{ \ba_j \in \mathbb{G}_f : ~ j = 1,2, \ldots, N_f \right\} \;  .
\ee
The union of all groups forms the whole society
\be
\label{3.91}
 \bigcup_f \; \mathbb{G}_f = \mathbb{G} \; , \qquad \sum_f N_f = N \; .
\ee

The spatial locations of groups is represented by the manifold indicator 
functions
\begin{eqnarray}
\label{3.92}
\xi_f(\ba_j) = \left\{ \begin{array}{ll}
1 \; , ~ & ~ \ba_j \in \mathbb{G}_f \\
0 \; , ~ & ~ \ba_j \not\in \mathbb{G}_f
\end{array} \right. \; .
\end{eqnarray}
The collection of all indicator functions, showing the society configuration, 
is denoted as
\be
\label{3.93}
\xi = 
\{ \xi_f(\ba_j): ~ \ba_j \in \mathbb{G} \; , ~ f = 1,2, \ldots \} \; .
\ee
The indicator functions possess the properties
\be
\label{3.94}
 \sum_f \xi_f(\ba_j) = 1 \; , \qquad \sum_j \xi_f(\ba_j) = N_f \; ,
\ee
the first of which means that each agent pertains to some of the groups, while 
the second shows the number of agents in an $f$-th group.   
       
In a realistic society, the agents can change their locations, so that the 
location vector ${\bf a}_j = {\bf a}_j(t)$ depends on time $t$. Hence the 
society configuration also is a function of time,
\be
\label{3.95}
\xi(t) = \{ \xi_f(\ba_j(t)) : ~ 
\ba_j(t) \in \mathbb{G} \; ; ~ f = 1,2, \ldots\} \; .
\ee
Since $t_{exp} \gg t_{int}$, the observable quantities describing the society 
correspond to the double average, over the system variables and over time, 
\be
\label{3.96}
\lgl \; A(\sgm,\xi(t)) \; \rgl = \sum_\sgm \frac{1}{t_{exp}} \;
\int_0^{t_{exp}} \rho(\sgm,\xi(t)) A(\sgm,\xi(t)) \; dt \; .
\ee

The motion of the groups inside the society usually is so much complicated 
that it is neither possible nor reasonable to follow the detailed movements 
of all agents, but the agent locations can be treated as random. Then it is 
possible to interchange the averaging over time with the averaging over the 
random society configurations by the rule
\be
\label{3.97}
\frac{1}{t_{exp}} \; \int_0^{t_{exp}} dt \longmapsto \int \cD \xi \;   .
\ee
Then we need to realize the functional integration over the manifold indicator 
functions. We will not describe the mathematical details of the integration 
that can be found in review articles \cite{Yukalov_36,Yukalov_37}, but will 
present the results.    

The probability distribution $\rho(\sigma,\xi)$ of a heterophase society, 
depending on the configuration $\xi$, can be derived following Sec. 2.2. The 
probability is to be normalized,
\be
\label{3.98} 
\sum_\sgm \int \rho(\sgm,\xi) \; \cD \xi = 1 \; .
\ee
The average energy is given by the expression
\be
\label{3.99}
 \sum_\sgm \int \rho(\sgm,\xi)H(\sgm,\xi) \; \cD \xi = E \; .
\ee
The information functional takes the form
$$
I[\; \rho(\sgm,\xi) \;] = \sum_\sgm 
\int \rho(\sgm,\xi)\; \ln\; \frac{\rho(\sgm,\xi)}{\rho_0(\sgm,\xi)} \; \cD \xi
\; +
$$
\be
\label{3.100}
 + \;
\lbd_0 \left[ \; \sum_\sgm \int \rho(\sgm,\xi) \; \cD\xi - 1 \; \right]
+ \lbd \left[ \; 
\sum_\sgm \int \rho(\sgm,\xi) \; H(\sgm,\xi) \; \cD\xi - E \; \right] \; .
\ee

Minimizing the information functional and assuming the uniformity of the trial 
distribution $\rho_0(\sigma,\xi) = const$ yields the probability
\be
\label{3.101}
\rho(\sgm,\xi) = \frac{1}{Z} \; e^{-\bt H(\sgm,\xi)} \;  ,
\ee 
with the partition function
\be
\label{3.102}
Z = \sum_\sgm \int e^{-\bt H(\sgm,\xi)} \; \cD\xi \; .
\ee

The free energy reads as
\be
\label{3.103}
F = - T \ln Z = - T \ln \sum_\sgm \int e^{-\bt H(\sgm,\xi)} \; \cD\xi \; .
\ee
Observable quantities are given by the averages
\be
\label{3.104}
\lgl \; A(\sgm,\xi) \; \rgl = 
\sum_\sgm \int \rho(\sgm,\xi) A(\sgm,\xi) \; \cD\xi \; .
\ee

Accomplishing the averaging over configurations allows us to define an 
effective Hamiltonian $\widetilde{H}$ by the relation
\be
\label{3.105}
e^{-\bt \widetilde H} = \int e^{-\bt H(\sgm,\xi)} \; \cD\xi \; .
\ee
The effective Hamiltonian takes the form
\be
\label{3.106}
\widetilde H = \bigoplus_f \; H_f \; , \qquad H_f = H_f(\sgm,w_f) \; ,
\ee
in which the probability that an agent pertains to a $j$-th group is
\be
\label{3.107}
w_f \equiv \frac{N_f}{N} \; .
\ee
In other words, this is the fraction of the society agents belonging to the 
$f$-th group. Of course, the normalization conditions are valid,
\be
\label{3.108}
 \sum_f w_f = 1 \; , \qquad 0 \leq w_f \leq 1 \; .
\ee
The group probabilities are the minimizers of the free energy
\be
\label{3.109}
 F = - T \ln \sum_\sgm e^{-\bt\widetilde H} \; ,
\ee
under the normalization conditions (\ref{3.108}),
\be
\label{3.110}
F ={\rm abs}\min F(w_1,w_2,\ldots) \; .
\ee

The observable quantities are given by the averages
\be
\label{3.111}
\lgl \; A(\sgm,\xi) \; \rgl = \sum_f \lgl \; A_f(\sgm,w_f) \; \rgl \; ,
\ee
where
\be
\label{3.112}
\lgl \; A_f(\sgm,w_f) \; \rgl = 
\sum_\sgm  \rho_f(\sgm,w_f) A_f(\sgm,w_f) \; ,
\ee
with the effective probability distributions 
\be
\label{3.113}
\rho_f(\sgm,w_f) = \frac{1}{Z_f} \; e^{-\bt H_f(\sgm,w_f)} \; ,   
\ee
and the partition functions
\be
\label{3.114}
 Z_f = \sum_\sgm e^{-\bt H_f(\sgm,w_f)} \; .
\ee

The sum of direct summation $\bigoplus$ in Eq. (\ref{3.106}) is used instead 
of the simple sum in order to stress that the summands are defined on different 
configuration sets
\be
\label{3.115}
X_f = \{ \sgm_{fj}:~ j = 1,2, \ldots , N_f \} \; ,
\ee
with the total society configuration set being the tensor product
\be
\label{3.116}
X = \bigotimes_f \; X_f \; .
\ee

This is the general approach for describing the heterophase societies with 
fluctuating groups. The approach reduces the consideration of quasi-equilibrium 
systems with mesoscopic group fluctuations to the description of effective 
equilibrium systems with a renormalized effective Hamiltonian. The mathematics 
of accomplishing functional integration over manifold indicator functions is 
described in reviews \cite{Yukalov_15,Yukalov_36,Yukalov_37}. More mathematical 
details are presented in Refs. \cite{Yukalov_38,Yukalov_39,Yukalov_40}. In the 
following section, we give an example of applying this approach.

\subsection{Self-Organized Disorder}

Let us consider the yes-no model of Sec. 3.5 modified so that to take account of 
fluctuating groups. For simplicity, we shall keep in mind two groups, one called 
ordered and the other disordered. The ordered group consists of agents agreeing with 
each other and characterized by a nonzero order parameter, while the disordered group 
is described by a zero order parameter to be defined below. 

Each $j$-th agent from an $f$-th group is characterized by the features $\sigma_{fj}$. 
In the yes-no model, these features are described by the variable $\sigma_{fj}=\pm 1$. 
The total feature set for the whole system is the configuration set (\ref{3.116}).

In addition to the interaction $J_{ij} >0$, typical for self-organization in the 
yes-no model, we need to include other interactions that would describe some 
disagreements between the agents. We denote the disordering interaction by $U_{ij}$. 

In the snapshot picture, the Hamiltonian with ordering and disordering interactions 
reads as
\be
\label{3.117}
 H(\sgm,\xi) = H_1(\sgm,\xi_1) \bigoplus H_2(\sgm,\xi_2) \; ,
\ee
with 
\be
\label{3.118}
H_f(\sgm,\xi_f) = 
\frac{1}{2} \sum_{i\neq j}^N \xi_f(\ba_i) \xi_f(\ba_j) U_{ij} - \; 
\frac{1}{2} 
\sum_{i\neq j}^N \xi_f(\ba_i) \xi_f(\ba_j) J_{ij} \sgm_{fi}\sgm_{fj} \; .
\ee
Here the manifold indicator functions show the belonging of the agents to the 
corresponding groups.

After averaging over group configurations, we obtain the effective Hamiltonian
$$
\widetilde H =H_1 \bigoplus H_2 \; , \qquad H_f = H_f(\sgm,w_f) \; ,
$$
\be
\label{3.119}
 H_f(\sgm,w_f) = \frac{1}{2} \; w_f^2 
\sum_{i\neq j}^N ( U_{ij} - J_{ij} \sgm_{fi}\sgm_{fj}) \;  .
\ee

The order parameter for the $f$-th group is
\be
\label{3.120}
s_f \equiv \frac{1}{N} \sum_{j=1}^N \lgl \; \sgm_{fj} \; \rgl \; ,
\ee
which translates into the equation
\be
\label{3.121}
 s_f = \tanh\left( \frac{J w_f^2 s_f}{T} \right) \; .
\ee
Let the first group be ordered, so that
\be
\label{3.122}
s_1 \neq 0 \;  ,
\ee
while the second group be disordered in the sense that
\be
\label{3.123}
 s_2 \equiv 0 \; .
\ee

In the case of two groups, it is convenient to denote the fraction of agents in the 
ordered group and disordered group as
\be
\label{3.124}
 w_1 \equiv w \; , \qquad w_2 = 1 - w \;  .
\ee
Then the necessary condition for the free-energy minimum is
\be
\label{3.125}
 \frac{\prt F}{\prt w} = 
\left\lgl \; \frac{\prt\widetilde H}{\prt w} \; \right\rgl = 0 \; .
\ee
This results in the equation
\be
\label{3.126}
w = \frac{u-s_2^2}{2u- s_1^2 - s_2^2} \;  ,
\ee
in which the notation is used:
\be
\label{3.127}
 u \equiv \frac{U}{J} \; , \qquad 
U \equiv \frac{1}{N} \sum_{i\neq j}^N U_{ij} \; ,
\qquad 
J \equiv \frac{1}{N} \sum_{i\neq j}^N J_{ij} \;  .
\ee
Taking into account that the disordered group is described by the zero order parameter
$s_2 = 0$, we have the fraction of agents in the ordered group
\be
\label{3.128}
 w = \frac{u}{2u-s_1^2} \; .
\ee
Due to the probability definition $0\leq w\leq 1$, and because $0 \leq s_1\leq 1$,
expression (\ref{3.128}) exists only for sufficiently strong disordering interactions, 
such that $u\geq 1$.  

Note that the second derivative of the free energy, to be positive, leads to the 
inequality
\be
\label{3.129}
\frac{\prt^2 F}{\prt w^2} =\left[ 
\left\lgl \frac{\prt^2\widetilde H}{\prt w^2} \right\rgl \; - \;
\bt \left\lgl \left( \frac{\prt\widetilde H}{\prt w} \right)^2 \right\rgl \right] ~ > ~ 0 \; ,
\ee
which yields the condition $u>1/2$. 
      
The society reduced energy 
\be
\label{3.130}
\overline E \equiv \frac{\lgl \widetilde H \rgl}{JN} 
\ee
leads to
\be
\label{3.131}
 \overline E = \overline E_1 + \overline E_2\; , \qquad 
E_f = \frac{1}{2} \; w_f^2 \left( u - s_f^2 \right) \; .
\ee

For simplicity, let us study the case of no noise, so that $T=0$. Then $s_1=1$ and 
the energy becomes
\be
\label{3.132}
 \overline E = 
w^2 \left( u - \; \frac{1}{2} \right) + \left( \frac{1}{2} - w \right) u \;  ,
\ee
while the fraction of ordered agents is 
\be
\label{3.133}
 w = \frac{u}{2u-1} \;  .
\ee
Thus the society energy is
\be
\label{3.134}
 \overline E = \frac{u(u-1)}{2(2u-1)} \; .
\ee

Now we have to decide what society is more stable, that which contains fluctuating 
disordered groups or that one which does not contain them? The more stable is that 
society whose free energy is lower. In the case of very small noise, we need to 
compare the corresponding expected energies. 

If no fluctuating groups would be allowed, then the ordered fraction would be exactly 
one $(w \equiv 1)$. The energy of a completely ordered society is 
\be
\label{3.135}
 E_{ord} \equiv \overline E(w=1) = \frac{u-1}{2} \;  ,
\ee
where $w \equiv 1$. If the whole society would contain only disordered agents 
$(w \equiv 0)$, then the society energy would be
\be
\label{3.136}
  E_{dis} \equiv \overline E(w=0) = \frac{1}{2} \; u \;  .
\ee
The society with disorder group fluctuations is more stable, when its energy is lower. 
Comparing the above expressions we see that
\be
\label{3.137}
 \overline E < E_{ord} < E_{dis} \qquad ( u > 1 ) \; .
\ee
Thus, for competing interactions, satisfying the condition $u > 1$, the energy of 
the society with fluctuating disorder is lower than that for the perfectly ordered 
society that, in turn, is lower than the disordered society. That is, a completely 
ordered society is more stable than a disordered society. However the most stable 
is a society with the admixture of fluctuating disorder. Thus disorder fluctuations 
make the society more stable. In that sense the disorder is self-organized. 

The total order parameter is defined as the average of the expression
\be
\label{3.138}
 M(\sgm,\xi) = \frac{1}{N} \sum_f \sum_{j=1}^N \xi_f(\ba_j) \sgm_{fj} \;  ,
\ee
which gives 
\be
\label{3.139}
 M = \lgl \; M(\sgm,\xi) \; \rgl = \sum_f w_f s_f \; .
\ee
When there is no noise, that is at zero temperature, we get
\be
\label{3.140}
 M = w s_1 = \frac{u}{2u-1} \qquad ( T = 0 ) \; .
\ee

Concluding, under sufficiently strong competition between the society agents, there 
spontaneously appear social disorder groups. When the disorder groups appear, they 
diminish the total society order. However, the existence of the disorder groups makes 
the society more stable. This is an example showing that it is not always useful to
try to realize a complete order in a society, since the presence of some disorder can
make the society more stable. Sometimes societies for being more stable generate 
self-organized disorder.

\subsection{Coexistence of Populations}

Very often social systems consist of several groups of rather different people,
so that the groups are more or less stable and on average do not essentially
change for long time, contrary to fluctuating groups considered in the previous 
section. Each of the groups can be characterized by different typical agents.
There exist numerous examples of such societies. Thus a society, forming a country,
very often can include groups having different nationality or religion.

In a society composing a financial market, there are the groups of fundamentalists 
and chartists. Fundamentalists base their decisions upon market fundamentals, such 
as interest rates, growth or decline of the economy, company's performance, etc.
Fundamentalists expect the asset prices to move towards their fundamental values, 
hence, they either buy or sell assets that are assumed to be undervalued or, 
respectively, overvalued. Chartists, or technical analysts, look for patterns and 
trends in the past market prices and base their decisions upon extrapolation of 
these patterns. There exist as well the groups of contrarians who buy or sell 
contrary to the trend followed by the majority.

Different groups of people or other alive beings, having differing features, 
are called populations. Let a society, inhabiting the volume $V$, contain 
several different populations enumerated by the index $n=1,2,\ldots$. In an 
$n$-th population, there are $N_n$ members. In what follows, we consider a 
general approach describing whether the populations can or cannot coexist with 
each other. The populations can be of any origin. For concreteness, we can keep 
in mind different population groups living in a country.

The populations can either peacefully live in the whole country, which can be 
called a mixed society, or can possess the tendency of separating from each 
other, thus having no intention for joint coexistence. Our aim is to understand 
how we could quantify the conditions when different populations prefer to live 
together and when they wish to 
separate. 

According to the general law, that society is more stable whose free energy is 
lower. That is, a mixed society, living in the same country, is more stable than 
the separated populations, existing separately from each other, if the free energy 
$F_{mix}$ of the mixed society is lower that the free energy $F_{sep}$ of the 
separated society,
\be
\label{3.141}
 F_{mix} < F_{sep} \;  .
\ee
The free energies can be represented as 
\be
\label{3.142}
F_{mix} = E_{mix} - T S_{mix} \; , \qquad
F_{sep} = E_{sep} - T S_{sep} \;   .
\ee
Therefore, denoting the entropy of mixing
\be
\label{3.143}
 \Dlt S_{mix} \equiv \frac{1}{N} \; ( S_{mix} - S_{sep} ) \;  ,
\ee
we have the condition of stability of the united country:
\be
\label{3.144}
 E_{mix} - E_{sep} < NT \Dlt S_{mix} \; .
\ee
     
We need to write down the energy of the mixed and separated populations. For this 
purpose, let us denote the density of an $n$-th population by $\rho_n({\bf r})$ 
and the interaction between the members of an $m$-th and $n$-th populations by 
$\Phi_{mn}({\bf r})$. Then the interaction energy for the mixed populations can 
be represented as
\be
\label{3.145}
E_{mix} = \frac{1}{2} \sum_{mn} 
\int_V \rho_m(\br) \; \Phi_{mn}(\br-\br') \; \rho_n(\br') \; d\br d\br' \; .
\ee
Assuming a uniform distribution of the people across the country implies the unform
densities
\be
\label{3.146}
\rho_m(\br) = \frac{N_m}{V} \; , \qquad  
\rho_n(\br) = \frac{N_n}{V} \;  .
\ee
Thus the interaction energy of a country with mixed populations reads as
\be
\label{3.147}
 E_{mix} = \frac{1}{2} \sum_{mn} \Phi_{mn} \; \frac{N_m N_n}{V} \;  ,
\ee
where the quantity
\be
\label{3.148}
\Phi_{mn} \equiv \int_V  \Phi_{mn}(\br)\; d\br 
\ee
describes the average interaction strength between the members of an $m$-th and 
$n$-th populations.

Now let us consider the case of separated populations, when each population lives 
in its separate location characterized by the volume $V_n$. Then the energy of a 
separated country is the sum
\be
\label{3.149}
E_{sep} = \frac{1}{2} \sum_n \int_{V_n} 
\rho_n^{sep}(\br)\; \Phi_{nn}(\br-\br')\; \rho_n^{sep}(\br')\; d\br d\br' \; .
\ee
Keeping again in mind that each population inside their part of the country is uniformly
distributed gives the uniform densities
\be
\label{3.150}
 \rho_n^{sep}(\br) = \frac{N_n}{V_n} \; .
\ee
Then the energy of a separated country is
\be
\label{3.151} 
 E_{sep} = \frac{1}{2} \sum_n \overline\Phi_{nn} \; \frac{N_n^2}{V_n} \; ,
\ee
where
\be
\label{3.152}
\overline\Phi_{nn} \equiv \int_{V_n} \Phi_{nn}(\br)\; d\br \;  .
\ee
Using the identity
$$
\frac{V}{V_m} = \sum_n \frac{V_m}{V_n}
$$
reduces the energy of a separated country to
\be 
\label{3.153}
 E_{sep} = 
\frac{1}{2} \sum_{mn} \overline\Phi_{mm} \; \frac{N_m^2N_n}{V V_m} \;  . 
\ee

Any two different populations, though being separated, exercise pressure on each 
other through their common boundary. If the pressure of one of them is larger than 
that of another, there is no equilibrium between the populations, but there can 
arise nonequilibrium movements, such as invasions and wars. The equilibrium 
coexistence of two separated populations implies the equality of their pressures:
\be
\label{3.154}
 P_m = P_n \qquad \left( P_n = - \; \frac{\prt F_{sep}}{\prt V_n} \right) \; .
\ee
By its meaning, this equality can be called {\it no-war condition}. For not too 
strong noise, such that 
$$
\frac{\prt F_{sep}}{\prt V_m} \; \cong \; \frac{\prt E_{sep}}{\prt V_m} =
\frac{1}{2} \; \overline\Phi_{mm} \left( \frac{N_m}{V_m} \right)^2 \;  ,
$$
we find the no-war condition in the form
\be
\label{3.155}
\frac{\overline\Phi_{mm}}{\overline\Phi_{nn}} = 
\left( \frac{N_n V_m}{N_m V_n} \right)^2 \; .
\ee
This shows that two neighboring populations have no war, being in equilibrium with 
each other, only when the signs of the effective interactions of the members inside 
each population are the same and the interactions satisfy the above conditions. An 
$m$-th population and $n$-th population can be in equilibrium, only when their internal 
interactions $\Phi_{mm}$ and $\Phi_{nn}$ are both either positive or negative. If for 
one population $\Phi_{mm}>0$, while for the members of another population $\Phi_{nn}<0$,
then there can be no equilibrium between such neighboring populations.
  
Taking account of the no-war condition (\ref{3.155}) makes it possible to rewrite the 
energy of the separated country in the form
\be
\label{3.156}
E_{sep} = \frac{1}{2} \sum_{mn} 
\frac{N_m N_n}{V} \; \sqrt{ \overline\Phi_{mm} \overline\Phi_{nn} } \;   .
\ee
Then the condition of stability (\ref{3.144}) for the mixed society yields the 
inequality
\be
\label{3.157}
 \sum_{mn} \frac{N_m N_n}{2V} \; \left( 
\Phi_{mn} - \; \sqrt{ \overline\Phi_{mm} \overline\Phi_{nn} } \; 
\right) ~ < ~  N T \Dlt S_{mix} \; .
\ee
With the identity
$$
N \equiv \frac{1}{N} \sum_{mn} N_m N_n \;   ,
$$
we find
\be
\label{3.158}
\sum_{mn} \; \frac{N_m N_n}{2V}  \left( 
\Phi_{mn} - \; \sqrt{ \overline\Phi_{mm} \overline\Phi_{nn} } \; - 
\; \frac{2T}{\rho} \; \Dlt S_{mix} 
\right) ~ < ~ 0 \; ,
\ee
where $\rho = N/V$ is the average density of the total population. From here, the 
sufficient condition of stability for the mixed country becomes
\be
\label{3.159}
 \Phi_{mn} ~ < ~ 
\sqrt{ \overline\Phi_{mm} \overline\Phi_{nn} } \; + \;
\frac{2T}{\rho} \; \Dlt S_{mix} \;  .
\ee
The entropy of mixture can be represented as
\be
\label{3.160}
 \Dlt S_{mix} = - \sum_m n_m \ln n_m \qquad 
\left( n_m \equiv \frac{N_m}{N} \right) \; .
\ee

Thus we obtain the condition of stability for the country with mixed population, as
compared to the country where the populations are separated,
\be
\label{3.161}
 \Phi_{mn} \; - \; \sqrt{ \overline\Phi_{mm} \overline\Phi_{nn} } ~ < ~
- \; \frac{2T}{\rho} \sum_m n_m \ln n_m \; .
\ee
When this condition is not valid, the country with mixed population is not stable and 
it will disintegrate into several separate countries, each being composed of just one 
population type. There exist numerous examples of countries that have disintegrated into
several smaller countries. This destiny, e.g., has happened with the Roman Empire, 
Makedonsky Empire, British Empire, French Empire, Austrian Empire, Ottoman Empire,
Soviet Union, Yugoslavia, and Czechoslovakia.

\subsection{Forced Coexistence}

From the history we know that empires are usually kept united not merely by economical
advantage of the countries composing them by also by force. Why this is possible and 
what happens when the required force becomes too strong? Is it possible to keep together 
different populations inside one empire by increasing force? To answer these questions,
it is necessary to consider the case with imposed regulations, including regulation cost.

Let the regulation force, acting on a member of an $n$-th population be $f_n(\br)$. 
The energy of a society with mixed populations, including the force applied for keeping 
different populations together, can be written as 
$$
E_{mix} = \frac{1}{2} \sum_{mn} 
\int_V \rho_m(\br) \; \Phi_{mn}(\br-\br') \; \rho_n(\br') \; d\br d\br' - 
\sum_n \int_V f_n(\br) \; \rho_n(\br) \; d\br +
$$
\be
\label{3.162}
 +
\frac{1}{2} \sum_{mn} \int_V \rho_m(\br) \; f_m (\br) \; A_{mn}(\br-\br') \; 
f_n(\br') \; \rho_n(\br') \; d\br d\br' \; ,
\ee
where the last term characterizes the cost of supporting the regulation force, with 
$A_{mn}({\bf r})$ being the strength of the interaction between the members of the 
society caused by the applied force at each other.

It is possible to assume that the population densities are uniform and the force is 
constant, so that
\be
\label{3.163}
 \rho_n(\br) = \frac{N_n}{V} \; , \qquad f_n(\br) = f_n \; .
\ee
Then the society energy reads as
\be
\label{3.164}
E_{mix} = \frac{1}{2} \sum_{mn} ( \Phi_{mn} + A_{mn} f_m f_n ) \; \frac{N_mN_n}{V} 
- \sum_n f_n N_n \; ,
\ee
where
\be
\label{3.165}
 A_{mn} \equiv \int_V A_{mn}(\br) \; d\br \; .
\ee

When the society is disintegrated into several independent countries, there is no 
need to apply force, hence the energy of a society with separated populations is 
(\ref{3.156}). Using the identity
$$
\sum_n f_n N_n = \frac{1}{2N} \sum_{mn} N_m N_n ( f_m + f_n )   
$$
and following the same way as in the previous section, we obtain the sufficient 
condition for the society with mixed populations
\be
\label{3.166}
\Phi_{mn} + A_{mn} f_m f_n ~ < ~ 
\sqrt{\overline\Phi_{mm} \overline\Phi_{nn}} \; + \; 
\frac{2}{\rho} \left( T \Dlt S_{mix} + \frac{f_m+f_n}{2} \right) \; .
\ee
For short, this condition can be named the condition of empire stability. If it 
does not hold, the empire disintegrates into separate countries. 

Condition (\ref{3.166}) shows that switching on the regulating force, first, 
increases the right-hand side of the inequality, thus making the mixed society 
more stable. However, too strong force makes the left-hand side of the inequality 
larger, thus leading to the instability of the empire and its disintegration. 
Hence, by enforcing reasonable regulations, it is possible to keep the coexistence 
of populations, even when without such an enforcement the country would disintegrate. 
However, the regulating force should not be too strong. This explains why, for some 
time, an empire can exist as a united family of different populations. But when the 
enforcement is too costly, it turns unbearable, and disintegration becomes inevitable. 
Empires do not exist forever.     

An empire, with forced coexistence of different populations, is practically always 
less stable than several independent countries, with their own populations and weaker 
regulation forces. However the unions of different populations also can exist, when 
they are kept not by force, but by their collaborative interactions. The prominent 
example is Switzerland, where the German, French, and Italian cantons peacefully 
coexist, not being forced, but due to the advantage of their mutual interactions.

\section{Collective Decisions in a Society}

The models considered above allow us to describe general structures and states of 
social systems, whose members exhibit specific behavior. Thus in the yes-no model,
the members are assumed to take decisions, either ``yes" or ``no", with regard to 
some problem, say voting in elections or buying something. As is evident, the 
underlying process of any action is the process of taking decisions. It is possible 
to say that practically all properties of a social system are caused by the decisions
taken by its members. This is why it is so important to have general understanding
of how decisions are made. In the present chapter, we consider some models 
describing the way of how the members of a society take decisions. Collective 
decisions are based on decisions made by individuals interacting with each other. 
Therefore, first of all it is necessary to understand how decisions are taken by 
individuals and then to model how they interact between themselves for reaching a 
collective decision. Moreover, there are deep parallels between the process of 
making decisions by individuals and by groups, since even individual decision making 
is a kind of collective decision making accomplished by the neurons of a brain. Since
the notions of decision making theory are less known for the physically oriented 
audience, we give below an overview of the main literature and describe the basic 
points of the theory.

\subsection{General Overview}

Nowadays, the dominant theory, describing individual behavior of decision makers 
is expected utility theory. This theory was introduced by Bernoulli \cite{Bernoulli_47},
when investigating the so-called St. Petersburg paradox. Von Neumann and Morgenstern 
\cite{Neumann_48} axiomatized this theory. Savage \cite{Savage_49} integrated within 
the theory the notion of subjective probability. The power of the theory was 
demonstrated by Arrow \cite{Arrow_50}, Pratt \cite{Pratt_51}, and by Rothschild and 
Stiglitz \cite{Rothschild_52,Rothschild_53} in the studies of risk aversion. The
flexibility of the theory for characterizing the attitudes of decision makers  
toward risk was illustrated by Friedman and Savage \cite{Friedman_54} and Markowitz 
\cite{Markovitz_55}. The expected utility theory has provided mathematical basis for 
several fields of economics, finance, and management, including the theory of games, 
the theory of investment, and the theory of search \cite{Lindgren_56,White_57,Rivett_58,
Berger_59,Marshall_60,Bather_61,French_62,Raiffa_63,Weirich_64,Gollier_65}. 

Despite many successful applications of the expected decision theory, quite a number 
of researchers have discovered a large body of evidence that decision makers, both 
human as well as animal, often do not obey prescriptions of the theory and depart 
from this theory in a rather systematic way \cite{Ariely_66}. Then there appear 
numerous publications, beginning with Allais \cite{Allais_67}, Edwards 
\cite{Edwards_68,Edwards_69}, and Ellsberg \cite{Ellsberg_70}, which experimentally 
confirmed systematic deviations from the prescriptions of expected utility theory, 
leading to a number of paradoxes. 

As examples of such paradoxes, it is possible to mention the Allais paradox 
\cite{Allais_67}, independence paradox \cite{Allais_67}, Ellsberg paradox 
\cite{Ellsberg_70}, Kahneman-Tversky paradox \cite{Kahneman_71}, Rabin paradox 
\cite{Rabin_72}, Ariely paradox \cite{Ariely_66}, disjunction effect \cite{Tversky_73},
conjunction fallacy \cite{Tversky_74,Shafir_75}, isolation effects \cite{Mccaffery_76}, 
planning paradox \cite{Kydland_77}, and dynamic inconsistency 
\cite{Strotz_78,Frederick_79}.

In order to avoid paradoxes, there have been many attempts to change the expected 
utility theory, which has been named as non-expected utility theories. There is a 
number of such non-expected utility theories, among which we mention just a few,
the most known approaches. These are prospect theory 
\cite{Edwards_68,Kahneman_71,Tversky_81}, weighted-utility theory 
\cite{Karmarkar_82,Karmarkar_83,Chew_84}, regret theory \cite{Loomes_85}, 
optimism-pessimism theory \cite{Hey_86}, dual-utility theory \cite{Yaari_87}, 
ordinal-independence theory \cite{Green_88}, quadratic-probability theory \cite{Chew_89}.
More detailed information on this topic can be found in the recent reviews by 
Camerer et al. \cite{Camerer_90} and Machina \cite{Machina_91}.   

Despite numerous attempts of treating expected utility theory, none of the suggested
modifications can explain all those paradoxes, as has been shown by Safra and Sigal 
\cite{Safra_92}. The best that could be achieved is a fitting for interpreting just 
one or a couple of paradoxes, with other paradoxes remaining unexplained. Moreover, 
spoiling the structure of expected utility theory results in the appearance of 
several inconsistences. Accomplishing a detailed analysis, Al-Najjar and Weinstein 
\cite{Al_93,Al_94} concluded that any variation of the expected utility theory 
``ends up creating more paradoxes and inconsistences than it resolves''.

An original interpretation was advanced by Bohr \cite{Bohr_95,Bohr_96,Bohr_97}, who 
suggested that psychological processes could be described by resorting to quantum
notions, such as interference and complimentarity. Von Neumann \cite{Neumann_98} 
mentioned that the theory of quantum measurements could be interpreted as decision 
theory. These ideas were developed in quantum decision theory \cite{Yukalov_99,Yukalov_100},
on the basis of which the classical decision-making paradoxes could be explained.
It has also been shown \cite{Yukalov_101,Yukalov_102} that quantum decision theory 
can be reformulated to the classical language employing no quantum formulas. 

When decision makers interact with each other, the process of decision making becomes
collective \cite{Krause_103,Seeley_104,Easley_105,Bourke_106,Marshall_107,Hamann_108}. 
Then each individual takes decisions being based not solely on personal deliberations, 
but also to some extent imitating the actions of other members of the society. 
Sometimes, this imitation grows to the level of herding effect. In the present chapter, 
we describe the main steps of how the process of decision making develops, 
starting from individual decisions to collective decisions taken by a network 
of society members.

\subsection{Utility Function}

The primary elements in a problem of choice are some events, or outcomes, or 
consequences, or payoffs denoted as $x$. One considers a set of outcomes, or 
a set of payoffs, or consumer set, or a field of events
\be
\label{4.1}
X = \{ x_i: i = 1,2,\ldots, N \} .
\ee
Generally, payoffs $x_i$ can be either finite or infinite, as well as the payoff 
set can be finite or infinite. Mathematically correct definition of infinity is 
through the limit of a sequence.
  
Payoffs or outcomes have to be measured in a common system of units, accomplished 
through a utility function $u(x)$, also called elementary utility function, pleasure 
function, satisfaction function, or profit function $u(x): \quad X \ra \mathbb{R}$.
The utility function has to satisfy the following properties:

\vskip 2mm

(i) It has to be nondecreasing, so that 
\be
\label{4.2}
u(x_1) \geq u(x_2) \qquad (x_1 \geq x_2 ) \; .
\ee
If (\ref{4.2}) is a strict inequality, the function is termed strictly increasing.

\vskip 2mm

(ii) It is often (although not always) taken to be concave, when
\be
\label{4.3}
u (\al_1 x_1 + \al_2 x_2 ) \geq \al_1 u(x_1) + \al_2 u(x_2),
\ee
where
$$
\al_1 \geq 0 \; , \qquad \al_2 \geq 0 \; , \qquad
\al_1 + \al_2 = 1 \; .
$$
It is called strictly concave, if (\ref{4.3}) is a strict inequality.

\vskip 2mm

A twice differentiable function is nondecreasing if
$$
u'(x) \equiv \frac{du(x)}{dt} \geq 0
$$
and it is concave when
$$
u''(x) \equiv \frac{d^2 u(x)}{dx^2} \leq 0 \; .
$$

The derivative $u'(x)$ defines the marginal utility function. According to the 
above properties, the marginal utility does not increase.

\vskip 2mm

An important property of a utility function is its {\it risk aversion}. The degree 
of absolute risk aversion is measured \cite{Pratt_51} by the quantity 
\be
\label{4.4}
r(x) \equiv -\; \frac{u''(x)}{u'(x)} \; .
\ee
Coefficient of relative risk aversion \cite{Arrow_50,Pratt_51} is defined as
\be
\label{4.5}
\widetilde r(x) \equiv x r(x) = - x \; \frac{u''(x)}{u'(x)}.
\ee

For a concave utility function, the degree of risk aversion is non-negative, hence 
utility function is risk averse, $r(x) \geq 0$. This means that with increasing $x$, 
the growth of the utility function $u(x)$ does not increase.

In practice, one often employs a {\it linear utility function}
\be
\label{4.6}
u(x) = k x  \qquad (k > 0 ) \; .
\ee
For this function,
$$
u'(x) = k \; , \qquad u''(x) = 0  \; ,
$$
hence the degree of risk aversion is zero, $r(x) = 0$.

Another used form is a {\it power-law utility function}
\be
\label{4.7}
u(x) = kx^\gm \qquad ( k > 0, \; 0 \leq \gm < 1) \; ,
\ee
assuming $x \geq 0$. For this case,
$$
u'(x) = \frac{k\gm}{x^{1-\gm}} \; , \qquad
u''(x) = - \frac{k\gm(1-\gm)}{x^{2-\gm}} \; .
$$
The degree of risk aversion diminishes with increasing $x$ as
$$
r(x) = \frac{1-\gm}{x} > 0 \; .
$$

One more example is the {\it logarithmic utility function}
\be
\label{4.8}
u(x) = c\ln ( 1 +kx ) \qquad ( c> 0,\; k>0) \;  ,
\ee
again keeping in mind $x \geq 0$. Then
$$
u'(x) = \frac{ck}{1+kx} \; , \qquad
u''(x) = -\; \frac{ck^2}{(1+kx)^2} \; .
$$
Hence the degree of risk aversion diminishes with the increasing payoff $x$, 
$$
r(x) = \frac{ck}{1+kx} \; .
$$

Sometimes, one uses an {\it exponential utility function}
\be
\label{4.9}
u(x) = c \left ( 1 - e^{-kx} \right ) \qquad (c>0, \; k>0) \; ,
\ee
for which
$$
u'(x) = cke^{-kx} \; , \qquad u''(x) = - ck^2 e^{-kx} \; ,
$$
and the degree of risk aversion is constant, $r(x) = k$.

In above examples, the utility function is twice-differentiable, such that $u'(x)$ 
and $u''(x)$ exist, it is non-decreasing, concave, hence risk averse, it is also 
non-negative for non-negative $x$, and it is normalized to zero, so that $u(0) = 0$,
implying that the utility of nothing is zero.

Some of the utility functions exemplified above are defined only for positive payoffs
$x>0$. In the case of losses, payoffs are negative. Then one defines different utility 
functions for gains and losses, say, $u_g(x)$ for gains and $u_l(x)$ for losses. Of 
course, these should coincide at zero, so that $u_g(0) = u_l(0)$.

It is also useful to mention that there are two types of utility, cardinal and ordinal.
Cardinal utility can be precisely measured and the magnitude of the measurement is 
meaningful, similarly how distance is measured in meters, or time in hours, or weight 
in kilograms. For ordinal utility, not its precise magnitude is important, but the
magnitude of the ratios between different utilities is only meaningful.

\subsection{Expected Utility} 

The notion of expected utility was introduced by Bernoulli \cite{Bernoulli_47} and 
axiomatic utility theory was formulated by von Neumann and Morgenstern \cite{Neumann_48}. 
The definitions below follow the classical exposition of von Neumann and Morgenstern 
\cite{Neumann_48}.

One defines a probability measure over the set of payoffs
\be
\label{4.10}
\{ p(x_i) \in [0,1]: \; i=1,2,\ldots,N\} \;   ,
\ee
with the probabilities $p(x_i)$ normalized to one,
\be
\label{4.11}
\sum_{i=1}^N p(x_i) = 1 \; .
\ee
The probabilities can be objective, prescribed by a rule \cite{Neumann_48}, or they 
can be subjective probabilities, evaluated by a decision maker \cite{Savage_49}.

The probability distribution over a set of payoffs is termed a {\it lottery},
\be
\label{4.12}
L = \{ x_i, p(x_i): \; i=1,2,\ldots,N \} \;   .
\ee

A compound lottery is a lottery whose outcomes are other lotteries. For two lotteries
$$
 L_1 = \{ x_i,p_1(x_i) \} \;  , \qquad
L_2 = \{ x_i,p_2(x_i) \} \;  ,
$$
the compound lottery is the linear combination
\be
\label{4.13}
\al_1 L_1 + \al_2 L_2 = \{ x_i,\; \al_1 p_1(x_i) + \al_2 p_2(x_i) \} \; ,
\ee
where
$$
\al_1 \geq 0 \; , \qquad \al_2 \geq 0 \; , \qquad
\al_1 + \al_2 = 1 \; .
$$

The {\it lottery mean} is the lottery expected value
\be
\label{4.14}
E(L) \equiv \sum_{i=1}^N x_i p(x_i) \;   .
\ee

{\it Lottery volatility}, lottery spread, or lottery dispersion is the variance
\be
\label{4.15}
{\rm var}(L) \equiv \sum_{i=1}^N x_i^2 p(x_i) - E^2(L) \; .
\ee
The lottery volatility or lottery dispersion is a measure of the lottery uncertainty.

{\it Expected utility} of a lottery is the functional
\be
\label{4.16}
U(L) \equiv \sum_{i=1}^N u(x_i) p(x_i) \; .
\ee
The expected utility is proportional to the lottery mean in the case of a linear 
utility function. Lotteries are ordered implying the following definitions. 

\vskip 2mm

Lotteries $L_1$ and $L_2$ are indifferent if and only if
\be
\label{4.17}
U(L_1) = U(L_2) \qquad (L_1 = L_2) \; .
\ee
A lottery $L_1$ is preferred to $L_2$ if and only if
\be
\label{4.18}
U(L_1) > U(L_2) \qquad (L_1 > L_2) \;   .
\ee
A lottery $L_1$ is preferred or indifferent to $L_2$ if and only if
\be
\label{4.19}
U(L_1) \geq U(L_2) \qquad (L_1 \geq L_2) \;   .
\ee

Expected utility satisfies the following properties.

\vskip 2mm

(i) {\it Completeness}. For any two lotteries $L_1$ and $L_2$, one of the conditions 
is valid:
\be
\label{4.20}
L_1 = L_2 \; , \quad L_1 < L_2 \; , \quad L_1 > L_2 \; ,
\quad L_1 \leq L_2 \; , \quad L_1 \geq L_2 \;   .
\ee

\vskip 2mm

(ii) {\it Transitivity}. For any three lotteries, such that $L_1 \leq L_2$ and 
$L_2 \leq L_3$, it follows that $L_1 \leq L_3$.

\vskip 2mm

(iii) {\it Continuity}. For any three lotteries, ordered so that $L_1 \leq L_2 \leq L_3$, 
there exists $\alpha \in [0,1]$ for which
\be
\label{4.21}
L_2 = \al L_1 + ( 1 - \al) L_3 \; .
\ee

\vskip 2mm

(iv) {\it Independence}. For any $L_1 \leq L_2$ and arbitrary $L_3$, there exists 
$\alpha \in [0,1]$, such that
\be
\label{4.22}
\al L_1 + (1 -\al) L_3 \leq \al L_2 + ( 1 - \al) L_3 \; .
\ee
These properties follow directly from the properties of the utility function described 
above.

\vskip 2mm

The standard decision-making process proceeds through the following steps. For the 
same set of payoffs $\{x_n: n = 1, 2,..., N\}$, there exist several lotteries
\be
\label{4.23}
L_n = \{ x_i,\; p_n(x_i) \} \qquad ( n = 1,2,\ldots ) \; .
\ee
Defining the utility function $u(x)$, one calculates the expected lottery utilities 
$U(L_n)$. Comparing the values $U(L_n)$, one selects the largest among them.

A lottery $L^*$ is named {\it optimal} when it is preferred to all others from the
given choice of lotteries. A lottery is optimal if and only if its expected utility 
is the largest:
\be
\label{4.24}
U(L^*) \equiv {\rm sup}_n U(L_n) \; .
\ee

The aim of a decision maker is assumed to choose an optimal lottery maximizing the 
related expected utility.

\subsection{Time Preference}

Sometimes people need to compare {\it present goods} that are available for use at 
the present time with {\it future goods} that are defined as present expectations 
of goods becoming available at some date in the future. {\it Time preference} is the 
insight that people prefer present goods to future goods 
\cite{Frederick_79,Samuelson_109}. 

Mathematical description of the time preference effect can be done as follows. Let 
us consider at time $t=0$ a lottery
\be
\label{4.25}
 L = \{ x_i , \; p(x_i) \} \qquad ( t = 0 ) \; .
\ee
With a utility function $u(x)$ at time $t=0$, the expected utility of the lottery is
\be
\label{4.26}
U(L) = \sum_i u(x_i) p(x_i) \qquad ( t = 0 ) \;  .
\ee
  
The lottery, expected at time $t>0$, has the form
\be
\label{4.27}
 L(t) = \{ x_i(t) , \; p(x_i(t)) \} \qquad ( t > 0 ) \;  .
\ee
Denoting the utility function at time $t>0$ as $u(x(t),t)$, we have the expected 
utility of $L(t)$ as
\be
\label{4.28}
U(L(t) ) = \sum_i u(x_i(t),t) p(x_i(t)) \; .
\ee

According to the meaning of time preference, the same goods at future time are 
valued lower then at the present time just because any goods can be used during 
the interval of time $[0,t]$. In the case of money, its value increases with 
time since it can bring additional profit through an interest rate, hence the 
utility of a fixed amount of money decreases with time. This can be formalized 
as the inequality
\be
\label{4.29}
u(x_i,t) < u(x_i) \qquad ( t > 0 ) \; .
\ee

It is possible to introduce a {\it discount function} $D(x,t)$ by the relation
\be
\label{4.30}
 u(x_i,t) = u(x_i) D(x_i,t) \; ,
\ee
with the evident condition
\be
\label{4.31}
 D(x_i,0) = 1 \; .
\ee
From the time preference condition (\ref{4.29}), we have
\be
\label{4.32}
 D(x_i,t) < 1 \qquad ( t > 0 ) \; .
\ee

In this way, if the payoffs at the present time $t=0$ and at a future time $t>0$
are the same, then the present lottery is preferable over the future one,
\be
\label{4.33}
 U(L(t) ) < U(L) \qquad ( x_i(t) = x_i ) \;  .
\ee

In particular, if the discount function is uniform with respect to the payoffs,
\be
\label{4.34}
 D(x_i,t) = D(t) \; ,
\ee
then 
\be
\label{4.35}
 U(L(t) ) = U(L) D(t) \; ,
\ee
where $D(0) = 1$. 

In decision making, one uses the following discount functions. 

\vskip 2mm

{\it Power-law discount function}
\be
\label{4.36}
D(t) = \frac{1}{(1+r)^t} \;   .
\ee

\vskip 2mm

{\it Exponential discount function}
\be
\label{4.37}
D(t) = \exp(-\gm t) \;   .
\ee  

Note that the power-law discount function (\ref{4.36}) is equivalent to the 
exponential form (\ref{4.37}), since they are related by the reparametrization
$$
1+r = e^\gm \; , \qquad \gm = \ln(1+r) \; .
$$

\vskip 2mm

{\it Hyperbolic discount function}
\be
\label{4.38}
D(t) =  \frac{1}{(1+t/\tau)^{\gm \tau}} \;   ,
\ee
with positive parameters $\gamma$ and $\tau$.

A detailed review on the effect of time preference is given in \cite{Frederick_79}.

\subsection{Stochastic Utility}

There exists a number of factors that influence decision making. These factors are 
random, or stochastic, because of which the related approach in decision making is 
named stochastic. These factors, for instance, are:

\vskip 2mm
(i) External conditions under which the decision is made, such as weather, situation 
in the country, relations with people, opinions of other people, ...

\vskip 2mm
(ii) Internal physical state of the decision maker, such as fatigue, illness, pain, 
influence of alcohol, ...

\vskip 2mm
(iii) Internal psychological state, including biases, emotions, mood, impatience, ...

\vskip 2mm
(iv) Previous experience, information, prejudices, religion, principles, character, ...

\vskip 2mm
(v) Occasional errors in deciding.

\vskip 2mm
Stochastic decision theory assumes that all these interrelated factors can be 
characterized by some variables, say $\xi$, called {\it states of nature}. The 
total collection of the states of nature forms the nature set $\{\xi\}$. The 
variables $\xi$ are random, stochastic. It is supposed that there should be 
given a probability measure $\mu(\xi)$ on the nature set $\{\xi\}$. Utility function
$u(x,\xi)$ becomes a random variable. Probability of payoffs $p(x,\xi)$ is also a 
random variable.

In that way, a lottery becomes a {\it stochastic lottery}
\be
\label{4.39}
L(\xi) = \{ x_i,\; p(x_i,\xi): \; \; i=1,2,\ldots,N \} \;   .
\ee
Respectively, we come to a {\it stochastic utility}
\be
\label{4.40}
U(L,\xi)  = \sum_{i=1}^N u(x_i,\xi) p(x_i,\xi) \; .
\ee
As far as the states on nature are random, one needs to average over these states, thus
coming to the expected utility
\be
\label{4.41}
U(L) \equiv \int  U(L,\xi) \; d\mu(\xi) \;  .
\ee

The process of stochastic decision making consists in the choice between several 
stochastic lotteries
\be
\label{4.42}
L_n(\xi) = \{ x_i,\; p_n(x_i,\xi): \; \; i=1,2,\ldots,N \} \; .
\ee
Technically, this implies the choice between several expected utilities
\be
\label{4.43}
U(L_n) = \int  U(L_n,\xi) \; d\mu(\xi) \; .
\ee
One needs to choose a lottery with the largest expected utility $U(L^*)$. This 
approach, however, confronts several serious difficulties.

\vskip 2mm
(i) It is not clear how the random variables should be incorporated into lotteries.

\vskip 2mm
(ii) Calculations become rather complicated.

\vskip 2mm
(iii) The nature set is not fixed, generally, depending on time.

\vskip 2mm
(iv) The nature set also can depend on the set of payoffs.

\vskip 2mm
(v) The explicit form of the nature states probability is not known and has to be 
postulated.

\vskip 2mm
More details on stochastic utility theory can be found, e.g., in the books,   
\cite{Pratt_110,Friedrich_111}.

\subsection{Affective Decisions}

People make decisions being based not merely on rational grounds, by calculating 
utility, but also being affected by emotions that are irrational. Several attempts 
have been made to take account of emotions in decision making by modifying expected 
utility, which is equivalent to some variants of non-expected utility models
\cite{Yaari_87,Reynaa_112,Bracha_113}. Here we present the main points of a 
probabilistic approach of taking into account emotions. First, this approach was 
formulated by resorting to techniques of quantum theory \cite{Yukalov_99,Yukalov_100},
however later it was shown \cite{Yukalov_101,Yukalov_102} that it can be reformulated 
in classical terms, without invoking any quantum expressions. The basics of the
{\it probabilistic affective decision theory} are as follows.      

The aim of any decision making is to choose an alternative from a set of several 
alternatives. Let the set of alternatives be denoted as 
\be
\label{4.44}
\mathbb{A} = \{ A_n : \; n = 1,2,\ldots , N_A \} \; .
\ee
Each alternative from this set is assumed to be equipped with a probability $p(A_n)$, 
with the normalization condition
\be
\label{4.45}
\sum_n p(A_n) = 1 \; , \qquad 0 \leq p(A_n) \leq 1 \; .
\ee
This probability shows how it is probable that the alternative $A_n$ can be chosen. 

An alternative $A_1$ is said to be stochastically preferable to $A_2$, if and only 
if
\be
\label{4.46}
 p(A_1 ) > p(A_2) \; .
\ee
Two alternatives are called stochastically indifferent, if and only if
\be
\label{4.47}
 p(A_1 ) = p(A_2) \;   .
\ee
An alternative $A_{opt}$ is stochastically optimal if its probability is maximal,
\be
\label{4.48}
 p(A_{opt} ) = \sup_n p(A_n) \;  .
\ee

The usefulness of an alternative is characterized by a {\it utility factor} $f(A_n)$,
whose form is to be prescribed by normative rules. This factor shows the probability
of choosing an alternative $A_n$ being based on rational understanding of its utility.
The standard probability normalization is applied:
\be
\label{4.49}
 \sum_n f(A_n) = 1 \; , \qquad 0 \leq f(A_n) \leq 1 \; .
\ee
One says that an alternative $A_1$ is more useful than $A_2$, if and only if
\be
\label{4.50}
 f(A_1 ) > f(A_2) \;   .
\ee
Two alternatives are equally useful, if and only if
\be
\label{4.51}
 f(A_1 ) = f(A_2) \;   .
\ee

Affective features of an alternative are represented by an {\it attraction factor}
$q(A_n)$, with the normalization
\be
\label{4.52}
  \sum_n q(A_n) = 0 \; , \qquad -1 \leq q(A_n) \leq 1 \;  .
\ee
Positive attraction factors imply attractive alternatives, while negative attraction 
factors mean that the corresponding alternatives are repulsive. An alternative $A_1$ 
is more attractive than $A_2$, if and only if 
\be
\label{4.53}
 q(A_1 ) > q(A_2) \;  .
\ee
Two alternatives are equally attractive, if and only if
\be
\label{4.54}
  q(A_1 ) = q(A_2) \;   .
\ee

The probability $p(A_n)$ of an alternative $A_n$ is a functional of the related
utility factor $f(A_n)$ and of attraction factor $q(A_n)$, such that the 
{\it rational boundary condition} be valid,
\be
\label{4.55}
p(A_n) = f(A_n) \; , \qquad  q(A_n) = 0 \; ,
\ee
when, in the absence of emotions, the probability of an alternative coincides with
the rational utility factor.  
 
The simplest form of the probability functional satisfying the rational boundary 
condition (\ref{4.55}) is prescribed by the {\it superposition axiom}
\be
\label{4.56}
 p(A_n) = f(A_n) + q(A_n) \; .
\ee

An explicit expression for the utility factor can be derived from the principle of 
minimal information of Sec. 2. Assume that for each alternative $A_n$ there exists
a utility functional $U(A_n)$, so that an average utility is given by the standard 
condition
\be
\label{4.57}
 \sum_n f(A_n) U(A_n) = U \;  .
\ee
Then the information functional, under conditions (\ref{4.49}) and (\ref{4.57}), 
reads as
\be
\label{4.58}
I[\; f(A_n) \; ] = \sum_n f(A_n) \; \ln \; \frac{f(A_n)}{f_0(A_n)}
+ \al \left[ \; 1 - \sum_n f(A_n) \; \right] + 
\bt \left[ \; U - \sum_n f(A_n) U(A_n) \; \right] \;  ,
\ee
where $f_0(A_n)$ is a trial distribution. Minimizing the information functional 
with respect to the utility factor $f(A_n)$, yields
\be
\label{4.59}
  f(A_n) = 
\frac{f_0(A_n) e^{\bt U(A_n)}}{\sum_n f_0(A_n) e^{\bt U(A_n)} } \; .
\ee
The trial distribution $f_0(A_n)$ can be taken following the Luce rule 
\cite{Luce_114,Luce_115,Gul_116}, 
\be
\label{4.60}
 f_0(A_n) = \frac{a_n}{\sum_n a_n} \;  ,
\ee
where $a_n$ is the attribute of the alternative $A_n$, having the form
\be
\label{4.61}
a_n = U(A_n) \; ,\qquad U(A_n) \geq 0
\ee
for semi-positive utility functionals and  
\be
\label{4.62}
a_n = \frac{1}{|\; U(A_n)\;|} \; , \qquad U(A_n) < 0
\ee
for negative utility functionals \cite{Yukalov_102,Yukalov_117,Yukalov_118}.

\subsection{Wisdom of Crowds}

{\it Wisdom of crowds} is the notion assuming that large groups of people are 
collectively smarter than individual members of the same groups. This concerns 
any kind of problem-solving and decision-making. The justification of this idea
is based on the understanding that the viewpoint of an individual can inherently 
be biased, being influenced by the individual emotions and prejudices, whereas 
taking the average knowledge of a crowd results in eliminating the noise of 
subjective biases and emotions, thus producing a more wise aggregated result 
\cite{Surowiecki_119,Sunstein_120,Page_121,Fiechter_122}. Talking about the wisdom 
of crowds, one keeps in mind the following crowd features: (i) The crowd should have 
a diversity of opinions. (ii) Each personal opinion should remain independent of 
those around them, not being influenced by anyone else. (ii) Each individual from 
the crowd should make their own decision based solely on their individual knowledge.
These conditions exclude the situation, when the crowd members consult with each 
other and mimic the actions of their neighbors, which can lead to herding effects. 
The latter will be considered in the next section.

Let us enumerate the members of a crowd, or a society, by the index $j=1,2,\ldots N$. 
According to the previous section, each member of the considered group chooses an 
alternative $A_n$ with the probability
\be
\label{4.63}
 p_j(A_n) = f_j(A_n) + q_j(A_n) \; ,
\ee
under the standard normalization condition 
$$
\sum_{n=1}^{N_A} p_j(A_n) = 1 \; , \qquad 0 \leq p_j(A_n) \leq 1 \; .
$$

The aggregate opinion implies the arithmetic averaging over the society members, which
yields the average probability
\be
\label{4.64}
p(A_n) \equiv \frac{1}{N} \sum_{j=1}^{N} p_j(A_n) \; ,
\ee
composed of the superposition of the average utility factor
\be
\label{4.65}
f(A_n) \equiv \frac{1}{N} \sum_{j=1}^{N} f_j(A_n) \; ,  
\ee
and the average attraction factor
\be
\label{4.66}
q(A_n) \equiv \frac{1}{N} \sum_{j=1}^{N} q_j(A_n) \; ,
\ee
thus coming to expression (\ref{4.56}). 

The utility factor is prescribed by a rational evaluation of the utility of the 
considered alternatives, hence weakly depending on subjective emotions. This means
that $f_j(A_n)$ is approximately the same for any group member, which is equivalent
to the condition
\be
\label{4.67}
 f_j(A_n) = f(A_n) \; .
\ee
Notice that the normalization conditions 
\be
\label{4.68}
\sum_{n=1}^{N_A} f_j(A_n) = 1 \; , \qquad 0 \leq f_j(A_n) \leq 1 
\ee
remain valid. 
  
The attraction factor, on the contrary, is subjective, being essentially influenced 
by the agent's emotions. In that sense, the attraction factor is a random quantity.
It is random because of several causes. First, its randomness is due to the evident 
fact of the variability of emotions experienced by quite different people. Second,
emotions of even the same person vary at different times. And third, emotions 
randomly influence the choice due to the generic variability and local instability 
of neural networks in the brain, as has been found in numerous psychological and 
neurophysiological studies \cite{Werner_123,Arieli_124,Gold_125,Glimcher_126,
Schumacher_127,Shadlen_128,Webb_129,Kurtz_130,Woodford_131}.       
 
Nevertheless, despite the intrinsic randomness of the attraction factor, some of 
its aggregate properties can be well defined. First of all, the {\it alternation law}
is satisfied:
\be
\label{4.69}
 \sum_{n=1}^{N_A} q_j(A_n) = 0 \;   .
\ee
This follows directly from equations (\ref{4.63}) and (\ref{4.68}). Expression
(\ref{4.63}) also tells us that the attraction factor for a $j$-th society member 
is in the range
\be
\label{4.70}
 - f_j(A_n) \leq q_j(A_n) \leq 1 - f_j(A_n) \;  .
\ee
Respectively, the aggregate attraction factor varies in the interval
\be
\label{4.71}
 - f(A_n) \leq q(A_n) \leq 1 - f(A_n) \;  .
\ee

The set of alternatives can be separated into three classes, depending on the signs 
of the attraction factors, the class of positive attraction factors, 
\be
\label{4.72}
 q_+(A_n) = q(A_n) > 0 \; ,
\ee
negative attraction factors, 
\be
\label{4.73}
  q_-(A_n) = q(A_n) < 0 \;  ,
\ee
and zero attraction factors, when $q(A_n)=0$. According to the limits (\ref{4.71}),
positive attraction factors are in the interval
 \be
\label{4.74}   
 0 < q_+(A_n) < 1 - f(A_n) \;  ,
\ee
while negative attraction factors are in the range
\be
\label{4.75}
 - f(A_n) < q_-(A_n) < 0 \; .
\ee

Non-informative priors for the attraction factors can be estimated by means of 
the related arithmetic averages. Recall that, if a quantity $y$ lays in an interval
$[a,b]$, its arithmetic average is $\overline{y} = (a+b)/2$. And if the interval 
limits are in the ranges $a_1 \leq a \leq a_2$ and $b_1 \leq b \leq b_2$, then they
are expressed through their averages, so that 
$$
\overline y = \frac{1}{2} \; \left( \; \overline a + \overline b \; \right) =
\frac{1}{2} \; \left(\; \frac{a_1+a_2}{2} + \frac{b_1+b_2}{2} \; \right) \; .
$$
Keeping in mind that $0 < q_+(A_n) < 1-f(A_n)$ and $-f(A_n) < q_-(A_n) < 0$, while 
$0 \leq f(A_n) \leq 1$, we obtain the {\it quarter law}
\be
\label{4.76}
  q_+(A_n) = \frac{1}{4} \; , \qquad q_-(A_n) = -\; \frac{1}{4} \; .
\ee
 
Employing the non-informative priors for the aggregate attraction factors, one can 
estimate the probability of alternatives, averaged over the crowd, as
\be
\label{4.77}
 p(A_n) = f(A_n) \; \pm \; \frac{1}{4} \; ,
\ee
depending on whether the alternatives on average are attractive or not. The quarter 
law has been found to be in very good agreement with empirical data 
\cite{Yukalov_102,Yukalov_132}.

\subsection{Herding Effect}

In the previous section, the process of decision making by a crowd of independent 
agents is considered. However, generally, the members of a society interact with 
each other, which can result in drastic changes in the agents behavior, such as 
the occurrence of herding effect \cite{Martin_133,Sherif_134,Smelser_135,Merton_136,
Turner_137,Hatfield_138,Brunnermeier_139}. There can exist two kinds of interactions
between the society members, which can lead to collective effects, such as herding.
First, the members can mimic the actions of others just replicating their behavior. 
And, second, the members communicate through information exchange. Strictly speaking, 
the correct description of collective effects, arising in the process of agent's 
interactions, requires to study temporal processes, simply because collective effects
need time to be formed and develop. The consideration of temporal collective effects 
is out of the scope of the present article. However, due to their importance for social 
systems, we delineate the principal points of how collective interactions can be 
incorporated into the affective decision theory. Details can be found in Refs.
\cite{Yukalov_140,Yukalov_141,Yukalov_142}. 

Multistep decision theory deals with quantities depending on time. Then the probability
of choosing by an agent $j$ an alternative $A_n$ at time $t$ is $p_j(A_n,t)$, with the 
normalization
\be
\label{4.78}
 \sum_{n=1}^{N_A} p_j(A_n,t) = 1 \; , \qquad
0 \leq p_j(A_n,t) \leq 1 \;  .
\ee
The utility factor becomes $f_j(A_n,t)$, with the normalization
\be
\label{4.79}
\sum_{n=1}^{N_A} f_j(A_n,t) = 1 \; , \qquad
0 \leq f_j(A_n,t) \leq 1 \;    ,
\ee
and the attraction factor reads as $q_j(A_n,t)$, satisfying the conditions
\be
\label{4.80}
\sum_{n=1}^{N_A} q_j(A_n,t) = 0 \; , \qquad
-1 \leq q_j(A_n,t) \leq 1 \;    .
\ee
 
Taking into account the tendency of society members to replicate the actions of 
others defines the probability dynamics  
\be
\label{4.81}
 p_j(A_n,t+1) = ( 1 - \ep_j) \left[\; f_j(A_n,t) + q_j(A_n,t) \; \right]
+ \frac{\ep_j}{N-1} \sum_{i(\neq j)}^N 
\left[\; f_i(A_n,t) + q_i(A_n,t) \; \right] \; ,
\ee
where $\varepsilon_j$ is a replication parameter satisfying the condition
\be
\label{4.82}
 0 \leq \ep_j \leq 1 \qquad ( j = 1,2, \ldots, N)  \; .
\ee

As is explained above, there are two types of agent interactions in a society, 
replication of the actions of other members and exchange of information. The 
latter is known to attenuate the influence of emotions, which results in the 
attenuation of the attraction factor. The attenuation of the emotion influence 
due to agent interactions is well confirmed by empirical observations 
\cite{Kuhberger_143,Charness_144,Blinder_145,Cooper_146,Sutter_147,Tsiporkova_148,
Charness_149,Charness_150,Chen_151,Liu_152,Charness_153,Yukalov_154}. The 
attenuation of the attraction factor is described
\cite{Yukalov_140,Yukalov_141,Yukalov_142,Yukalov_154} by the form
\be
\label{4.83}
q_j(A_n,t) = q_j(A_n) \exp \{ - M_j(t) \} \;   ,
\ee
where $q_j(A_n)$ is the attraction factor of an agent $j$ in the absence of social 
interactions, $M_j(t)$ is the amount of information received by the moment of time 
$t$ by an agent $j$. The quantity $M_j(t)$ that, for short, can be called 
{\it memory}, writes as
\be
\label{4.84}
 M_j(t) = \sum_{t'=0}^t \; \sum_{i=1}^N
J_{ji}(t,t') \mu_{ji}(t') \;  .
\ee
Here $J_{ji}(t,t')$ is the information transfer function from an agent $i$ to the 
agent $j$ in the interval of time from $t'$ to $t$ and $\mu_{ji}(t)$ is the 
Kullback-Leibler information gain, received by the agent $j$ from the agent $i$ at 
time $t$,
\be
\label{4.85}
 \mu_{ji}(t) = \sum_{n=1}^{N_A} p_j(A_n,t) \; \ln \;
\frac{p_j(A_n,t)}{p_i(A_n,t)} \;  .
\ee

Interaction of the society members through replication and information exchange 
results in a rich variety of behavior types, including herding, periodic cycles,
and chaotic fluctuations, which depends on the society parameters. Detailed analysis 
of multistep decision making by intelligent members of a society is given in 
Refs. \cite{Yukalov_140,Yukalov_141,Yukalov_142}.

\section{Conclusion}

In the present part of the lectures, the principle of minimal information is
formulated, which gives the key for constructing probability distributions for
equilibrium and quasi-equilibrium social systems. Several simple examples, based 
on the yes-no model, are considered. Despite their seeming simplicity, the models 
allow one to describe rather nontrivial effects including the role of regulation cost
and the existence of fluctuating groups inside a society. Quite surprisingly, it turns 
out that the occurrence of self-organized disorder can make a society more stable.
The peculiarity of coexisting populations explains when these populations can mix
and peacefully live in the same country and when this coexistence becomes unstable, 
so that the country separates into several pieces, with different populations in
different locations. Since behind all actions of any society there are decisions
of the society members, the basics of the probabilistic affective decision theory
are delineated.      

Hopefully, the material of the above survey gives the reader a feeling of wide
possibility of applying mathematical models of physics for describing social 
systems, even being limited to equilibrium systems. The presented content is 
based on the lectures that have been given by the author during several years 
at the Swiss Federal Institute of Technology in Z\"{u}rich (ETH Z\"{u}rich). 
Being restricted by lecture bounds, it is, certainly, impossible to present 
numerous existing approaches and models of social systems, which can be found 
in the cited literature. This is why the topics touched upon are by necessity 
limited. The choice of the material is motivated by the interests of the author, 
which, not surprisingly, are often connected with the development of the theories 
and models he has been involved in.

The following part of the lectures will be devoted to nonequilibrium systems 
studying various evolution equations and dynamical effects.   

\vskip 2mm

{\bf Funding}: This research received no external funding.

\vskip 2mm

{\bf Acknowledgments}: I appreciate very much hospitality of and numerous 
discussions with D. Sornette without whose influence this work would not have 
been done. I am grateful for discussions and help to E.P. Yukalova.    
  
\vskip 2mm

{\bf Conflicts of Interest}: The author declares no conflict of interest.

\end{document}